\documentclass[preprint,11pt]{elsarticle}

\usepackage{amssymb}
\usepackage{float}
\usepackage{geometry}
\usepackage[caption = false]{subfig}
\usepackage{graphicx}
\usepackage{amsmath}
\usepackage{xcolor}
\usepackage{lineno}
\usepackage{multirow}
\usepackage{amsthm}
\usepackage{amsmath}
\usepackage{esvect}
\usepackage{hyperref}

\newcommand{\norm}[1]{\left\lVert#1\right\rVert}

\newtheorem{assumption}{Assumption}

\journal{Chaos, Solitons \& Fractals}

\begin{document}

\newgeometry{left=2.2cm,right=2.2cm, bottom=2.5cm, top=2.5cm}
\begin{frontmatter}



\title{Efficient and faithful reconstruction of dynamical attractors\\ using homogeneous differentiators} %Having fun with Differentiators}

\author[inst1]{Uros Sutulovic}

\affiliation[inst1]{organization={Department of Industrial Engineering, University of Trento},%Department and Organization
            addressline={via Sommarive 9}, 
            city={Trento},
            postcode={38123}, 
            country={Italy}}

\author[inst1]{Daniele Proverbio}
\author[inst1]{Rami Katz}
\author[inst1]{Giulia Giordano}

\begin{abstract} % max 250 words
Reconstructing the attractors of complex nonlinear dynamical systems from available measurements is key to analyse and predict their time evolution.
Existing attractor reconstruction methods typically rely on topological embedding and may produce poor reconstructions, which differ significantly from the actual attractor, because measurements are corrupted by noise and often available only for some of the state variables and/or their combinations, and the time series are often relatively short. Here, we propose the use of Homogeneous Differentiators (HD) to effectively de-noise measurements and more faithfully reconstruct attractors of nonlinear systems. Homogeneous Differentiators are supported by rigorous theoretical guarantees about their de-noising capabilities, and their results can be fruitfully combined with time-delay embedding, differential embedding and functional observability.
We apply our proposed HD-based methodology to simulated dynamical models of increasing complexity, from the Lorenz system to the Hindmarsh-Rose model and the Epileptor model for neural dynamics, as well as to empirical data of EEG recordings. In the presence of corrupting noise of various types, we obtain drastically improved quality and resolution of the reconstructed attractors, as well as significantly reduced computational time, which can be orders of magnitude lower than that of alternative methods. Our tests show the flexibility and effectiveness of Homogeneous Differentiators and suggest that they can become the tool of choice for preprocessing noisy signals and reconstructing attractors of highly nonlinear dynamical systems from both theoretical models and real data.
\end{abstract}

%Given available prescribed measurements, existing attractor reconstruction methods for systems described by an underlying dynamical model typically rely on topological embedding.
%Homogeneous Differentiators are based on tools from sliding mode control, homogeneous systems and differential inclusions, and their results can be fruitfully combined with time-delay embedding, differential embedding and functional observability. Differently from existing methods, HD are accompanied by rigorous theoretical guarantees about their de-noising capabilities.


    % Submit highlights as a separate editable file in the online submission system with the word "highlights" included in the file name.

    % Highlights should consist of 3 to 5 bullet points, each a maximum of 85 characters, including spaces.

\begin{highlights}
\item We reconstruct dynamical attractors of complex nonlinear systems from noisy data
\item We use Homogeneous Differentiators for efficient and faithful reconstruction
\item The results are more accurate and much faster than with alternative methods
\item The approach works successfully with various noise types and complex models
\item We also reconstruct de-noised attractors from empirical data of neural activity
\end{highlights}


\end{frontmatter}

%\linenumbers

\section{Introduction, Literature Overview and Motivation}
\label{sec:sample1}

Recent literature has shown an increasing interest in the analysis of complex systems and of their properties, effectively integrated with data analysis \cite{bianconi2023complex,artime2024robustness}; of particular importance is the extraction of system-level features from empirical data, for example to identify early warning signals and devise adaptive responses \cite{artime2024robustness}.
%To quantitatively analyse, understand and forecast the time evolution of complex dynamical systems, reconstructing their dynamical attractors is a key step, which is crucial by itself and also enables subsequent studies based, \textit{e.g.}, on topological data analysis \cite{patania2017topological,xu2021topological}. 
Reconstructing the dynamical attractors of complex dynamical systems from empirical data is a key step to quantitatively analyse, understand and forecast their time evolution, as well as to enable subsequent studies based, \textit{e.g.}, on topological data analysis \cite{patania2017topological,xu2021topological}. 
Attractor reconstruction sheds light onto fundamental topological properties of nonlinear systems \cite{sauer1991embedology,Marwan2009,Bhat2022}, with applications ranging from the characterisation of chaotic dynamics \cite{kennel1992method,Marwan2007} to the detection of causality and patterns in biological, ecological and economic data \cite{ahamed2021capturing,Sugihara2012,Groth2017}, from the diagnosis of diseases \cite{Carvalho2018, PerezToro2020} to the identification of geometric patterns and features of brain functions \cite{lucas2024topological,pourdavood2024eeg}.

Several methods have been developed to reconstruct attractors from measured time series, assuming an underlying dynamical model.
%and to effectively integrate empirical data and the analysis of complex system properties

Measured time series data are often relatively short, which poses a challenge to reconstruction algorithms, and they are typically available only for some of the state variables and/or combinations thereof. When the system is \textit{observable}, namely, its entire state can be fully reconstructed from the available measurements, recorded time series can be mapped into an attractor in the system's state space \cite{Letellier2005}; in the case of \textit{functional observability}, the reconstruction can be achieved in the state space for functions of the state variables \cite{montanari2022functional}. Even in the absence of observability, or functional observability, attractor reconstruction can be performed through embedding methods, such as \textit{time-delay embedding} \cite{Takens1981,Stark1997}, that map measurement data into a different space that is topologically equivalent to the original one up to a homeomorphism.
Moreover, measured data are unavoidably subject to corruption due to the presence of noise, which is a ubiquitous phenomenon in any realistic measurement process: often, this significantly degrades the quality of the measured signal and impedes the extraction of faithful information related to the underlying (clean) signal \cite{tuzlukov2018signal}, thus making the application of attractor reconstruction methods extremely challenging in real-world applications.
Even for studies aimed at extracting topological information \cite{patania2017topological, xu2021topological} such as homology metrics \cite{atienza2016separating, expert2019topological}, which are often conducted on noisy attractors \cite{tauzin2020giottotda}, noise of various nature may compromise the %(quantitative) % If you miscount the number of holes, it is more than just quantitative...
accuracy of the results, and adequate pre-processing steps for de-noising are thus required.

Filtering theory \cite{shenoi2005introduction}, developed to address the need to faithfully extract and process signals in high signal-to-noise-ratio (SNR) regimes, offers several methods to de-noise signals, often based on Fourier analysis (associating noise with certain decomposed frequency bands, usually high-frequency, and then applying an appropriate low-pass or band-pass filter); these methods can be used for attractor reconstruction from noisy measurements.
% An alternative modelling approach for signal estimation in the presence of noise employs a stochastic formulation in which the signal is a stochastic process, given as a solution to an Itō stochastic differential equation, and is reconstructed from noisy observations by seeking a \textit{causal} estimator (only depending on past observations) that is the closest possible in expectation to the original signal \cite{jazwinski2007stochastic,bain2009fundamentals}.
Still, noise particularly hinders attractor reconstruction in the original system's state space, as this requires knowledge not only of the (de-noised) measured signal, which can be recovered through filtering, but also of its (de-noised) derivatives. The differentiation problem \cite{levant2003higher,levant2017sliding,levant2020robust,hanan2021low} is precisely a generalisation of the filtering problem aimed at recovering not only a de-noised estimate of the signal, but also of its derivatives: it is therefore much more challenging and fundamentally ill-posed. In fact, noise can never be fully filtered out in practice and, as a consequence, taking derivatives of the signal (noisy, even after filtering) implies also differentiating the noise itself. Noise differentiation introduces potentially unbounded errors that yield an arbitrarily large decrease in SNR when estimating the signal derivatives, and thus an arbitrarily large corruption of the derivative estimates, even when the noise is very small.
Nevertheless, the availability of well-cleaned and de-noised signals \emph{along with their derivatives} is crucial for numerous problems and algorithms in applied mathematics, engineering and the life sciences, including data and signal processing \cite{woltring1985optimal}, radar theory \cite{mahalik2019estimates} and, as we show in this manuscript, attractor reconstruction.

Two main approaches have emerged to reconstruct attractors from noisy measurements, as a more sophisticated alternative to \textit{direct embedding} (which consists of directly applying the embedding methods \cite{Takens1981,Stark1997} to the noisy measured signal).
The first approach relies on an initial step to de-noise the measured signals and then reconstructs the attractor from the de-noised signals using classic embedding methods \cite{Takens1981,Stark1997}; signal de-noising approaches adopted so far to enable a subsequent attractor reconstruction include simple low-pass filtering and soft-thresholding of time-series \cite{porporato2001multivariate, wang2016application}, as well as linear approaches such as autoregressive moving average (ARMA) \cite{jayawardena2000noise}.
The second approach performs classic attractor reconstruction \cite{Takens1981,Stark1997} based on the noisy signals (direct embedding) and then applies, to the reconstructed noisy attractor, manifold de-noising techniques that typically rely on the Schreiber-Grassberger algorithm \cite{SchreiberGrassberger1991,grassberger1993noise, hegger1999practical}. The Schreiber-Grassberger algorithm offers improved noise reduction with respect to low-pass filters applied to the measurements before attractor reconstruction \cite[Section V.C]{grassberger1993noise}, and is still the most frequently used technique across fields \cite{sun2024aerial,heltberg2021tale,acevedo2021self,kiran2021nonlinear}.
For each point of the sequence in the time-delay coordinates, the algorithm uses the neighbouring past and future points to fit a local linear model via a least-squares estimate, and then it projects the point along the leading eigenvector of the resulting covariance matrix, so as to obtain a de-noised point aimed at being closer to the noise-free system trajectory.
%It corrects each point in the time-delay coordinates along a certain eigendirection of a covariance matrix with ``intensity'' proportional to a weighted average of neighbouring points
%At each time-point, it uses past and future values to fit a local linear model via a least-square estimate, and then it projects the point onto this model to obtain a new point that is closer to the noise-free system trajectory.
%An alternative approach for attractor de-noising applies a point-wise correction using a non-parametric Bayesian approach \cite{sk2016denoising}, but increases the computation steps and time and has not been widely adopted.

%Common limitations for these various approaches are that they
The above approaches have been primarily designed for low and bounded (and often additive Gaussian) noise \cite{muldoon1998delay, schouten1994estimation}, have been seldom tested on noise types that are commonly observed in measured time-series from natural systems, such as harmonic, unbounded, or multiplicative noise \cite{schreiber1999interdisciplinary, cheng2015time, haddad2010neuroadaptive, rudolph2003characterization}, have limited theoretical guarantees about their range of applicability and de-noising capabilities, and require long computational times.
Developing methods to efficiently and faithfully reconstruct attractors from noisy measurements is therefore still an open challenge.

In this work, we address the problem of reconstructing attractors of nonlinear dynamical systems from partial and noisy measurements by applying, for the first time, the theory of Homogeneous filtering Differentiators (from here on simply Differentiators, or HDs) introduced in \cite{levant2003higher,levant2017sliding,levant2020robust,hanan2021low}. 
Given a smooth signal that is corrupted by (possibly) non-smooth noise, HDs extract not only a cleaned estimate of the base signal, but also estimates of  its derivatives up to a prescribed order: since a HD can successfully de-noise corrupted measurement signals along with their derivatives,
we show that it can faithfully and accurately reconstruct system attractors \emph{also in the original state space}.
In particular, the HD is a nonlinear dynamical system whose state variables converge \textit{in finite time} to estimates of the derivatives of a scalar input signal. It was developed within the field of Sliding-Mode Control (SMC), also relying on tools from homogeneous systems and differential inclusions, to help stabilise systems affected by large uncertainty by means of a non-smooth control action.
The use of Homogeneous Differentiators, hitherto unexplored in attractor reconstruction, brings forth multiple significant advantages over existing methods, as we demonstrate in this manuscript.
First, the HD relies on SMC theory, which requires only very basic assumptions on the noise properties and thereby allows to theoretically cope with a large class of corrupting noises; this feature gives the HD the flexibility to handle a variety of de-noising scenarios, as opposed to being tailored only to a specific setting, such as Gaussian white noise.
Second, the HD provides \emph{finite-time} de-noising: its internal states converge to the desired estimates after a finite time, rather than asymptotically.
Third, rigorous theoretical guarantees can be proven for the estimation accuracy when the noise is additive: 
the HD offers the best theoretically possible de-noising guarantees (up to a universal multiplicative constant) with respect to the noise magnitude, namely, the estimates generated with the HD achieve the optimal theoretical error bounds \cite{levant2020robust}.
Fourth, its setup requires minimal knowledge of the system under study: the HD has few parameters that are easy to tune without compromising performance.
Finally, the HD enjoys significantly improved computational efficiency, with the computation performed in real time, in comparison to other state-of-the-art de-noising methods; specifically, here we compare it to the widely used Schreiber-Grassberger algorithm \cite{SchreiberGrassberger1991,grassberger1993noise} (a comparison to de-noising via a Kalman filter can be found in \cite{levant2020robust}).
The above properties are essential for realistic on-line applications that require rapid de-noising of the corrupted signal, such as early warnings of medical states \cite{yamanashi2021topological, gavidia2024early}.

We systematically investigate the use of the Differentiator to reconstruct dynamical attractors from measured noisy and partial time series. We show that it is a powerful tool that achieves simultaneously both accurate de-noising of the measured signal, even when the underlying model is unknown, and faithful attractor reconstruction directly in the original state-space coordinates, when the system is known to be observable, because it also provides estimates of the time derivatives of the de-noised measured signal (see Figure~\ref{fig:Differentiator_advantages}). The HD achieves a faithful reconstruction with low computational time complexity, for all the tested noise types and intensities, both in the additive and in the multiplicative case; for additive noise, rigorous theoretical guarantees can be provided on the estimation error bounds.

We thoroughly review the theory of HDs in Section~\ref{sec:methods} (as well as in Section S1 of the Supplementary Material), where we also discuss their applicability for attractor reconstruction from time series and provide indications for their computational implementation and practical usage.
In Section~\ref{sec:results}, we conduct our numerical investigation on models of increasing complexity, subject to a variety of noise types, both additive and multiplicative. In Section~\ref{sec:Lorenz}, we study the benchmark example of the chaotic Lorenz'63 system \cite{Lorenz63} and we quantify the performance of the HD compared to a commonly employed noise-reducing method, in terms of quality of the attractor reconstruction and required computational effort. In Section~\ref{sec:Hindmarsh-Rose}, we consider the Hindmarsh-Rose (HR) model \cite{hindmarsh1984model} for neural dynamics and use the attractor reconstruction results obtained from applying the HD on measurements of a single variable to study neural computation properties without access to direct measurements of the membrane potential. In Section~\ref{sec:Epileptor}, for the Epileptor model \cite{jirsa2014nature} (an extension of the HR model for simulation of seizure-like events in groups of neurons), we investigate how applying the HD on measurements of multiple variables improves the quality of attractor reconstruction in a large-scale system.
Finally, in Section~\ref{sec:empirical_data}, we consider real measurement data and apply the HD to reconstruct the dynamic attractor of epileptic brain activity from empirical EEG signals; our results support the use of the HD to preprocess noisy signals and reconstruct attractors of highly nonlinear dynamical systems also based on real noisy data.
Section~\ref{sec:discussion} provides a final discussion of the results and of possible future directions.

\section{Methods: leveraging Homogeneous Differentiators for Attractor Reconstruction}
\label{sec:methods}

\subsection{The Homogeneous filtering Differentiator (HD) in continuous and discrete time}
\label{subsec:hom_filt_diff}

In this section, we describe the basic properties of Homogeneous Differentiators \cite{levant2003higher,levant2017sliding,levant2020robust,hanan2021low}; for a more detailed exposition and a formal presentation of the theoretical de-noising guarantees offered by HDs, see Section S1 in the Supplementary Material.

Consider a scalar measured signal of the form $f(t) = f_0(t)+\eta(t)$, $t\in \mathbb{R}_+$, where $f_0(t)$ is a (clean) base signal that is corrupted by a noise $\eta(t)$. Let us denote by $f^{(i)}_0(t)$ the $i$-th time derivative of $f_0(t)$. For a pre-defined number $n_d\in \mathbb{N}_0$, the differentiation problem consists of estimating $f^{(0)}_0(t),\dots, f^{(n_d)}_0(t)$ in \emph{finite time} from the measured signal $f(t)$. We make the following assumptions:

\begin{assumption}\label{assump:BaseSignal}
The base signal $f_0(t)$ is $n_d+1$ times continuously differentiable. Moreover, there exists a \emph{known} constant $L_0>0$ such that $\left| f_0^{(n_d+1)}(t)\right|\leq L_0$ for all $t\in \mathbb{R}_+$.
\end{assumption}

\begin{assumption}\label{assump:Noise}
The noise $\eta(t)$ has $n_f+1\in \mathbb{N}$ components, $\eta(t) = \sum_{j=0}^{n_f}\eta_j(t)$, with $n_f \in \mathbb{N}_0$. The component $\eta_j(t), \ j=0,\dots, n_f$ is of the global filtering order $j$ with $j$-th order global integral magnitude ${\epsilon}_j \geq 0$ (see Definition 1, Section S1 in the Supplementary Material). While $\eta_1,\dots, \eta_{n_f}$ can be unbounded, $\eta_0$ is a measurable and essentially bounded function.
\end{assumption}
Assumption~\ref{assump:BaseSignal} relates to the smoothness properties of the base signal $f_0(t)$, whose derivatives one wants to recover. Assumption~\ref{assump:Noise} on the noise is needed in order to derive rigorous theoretical estimates on the differentiation errors, but cannot be verified in practice, as the noise is assumed to be unknown; hence, the value of the filtration degree $n_f$ is chosen from physical considerations about the source of the noise, on the one hand, and computational considerations (complexity), on the other hand.
Subject to Assumptions~\ref{assump:BaseSignal} and~\ref{assump:Noise}, the continuous-time homogeneous filtering differentiator scheme has the following two-block ordinary differential equation (ODE) structure
\begin{equation}\label{eq:DifferentiatorStruct}
\begin{array}{llll}
\hspace{-5mm} & \begin{cases}
w_1(t) = z_0(t)-f(t),\, & \mbox{ for }  n_f=0,\\
\dot{w}(t)= \Omega_{n_d,n_f}(w(t),z_0(t)-f(t),L_0,{\lambda}),\, & \mbox{ for } n_f > 0, 
\end{cases}
& \, \mbox{ and } \,\, \dot{z}(t)=D_{n_d,n_f}(w_1(t),z(t),L_0,{\lambda}),
\end{array}
\end{equation}
where $w = (w_1,\dots, w_{n_f})^{\top}$ is the state of the filtration block, $z=(z_0,\dots,z_{n_d})^{\top}$ is the state of the differentiation block, $f$ is the measured (noisy) signal and $\Omega_{n_d,n_f}$, $D_{n_d,n_f}$ are appropriate vector-valued functions (whose complete expression is reported above Eq. (S1) in the Supplementary Material). The scheme is initiated with the parameters $n_d$, $n_f$ and $L_0$ supplied by the user. The additional parameters in vector $ \lambda =(\tilde{\lambda}_0,\dots, \tilde{\lambda}_{n_d+n_f})$ are extracted from a pre-specified array and can be explicitly provided for $n_d+n_f\leq 12$ (however, a suitable set of parameters is proven to exist for all possible $n_d$ and $n_f$, via a recursive construction \cite{levant2003higher}; see Section S1.1 in the Supplementary Material). Intuitively speaking, the $w$-system (filtration block) is responsible for eliminating the noise components $\eta_1,\dots,\eta_{n_f}$ by filtering them out of the signal $f$, so that in finite time $w_1(t)\approx f_0(t)+\eta_0(t)$ holds. Then, $w_1$ is fed into the $z$-system, which performs the differentiation and produces estimates $z_i(t)\approx f_0^{(i)}(t),\ i=0,\dots,n_d$ in finite time and without resorting to differentiation of the noise components. An important feature of the HD, crucial since the true nature of the corrupting noise $\eta(t)$ is not known, is that increasing the filtration degree $n_{f}$ does not deteriorate the performance; by taking larger values of $n_f$, one can filter out more complex noises and potentially reduce the differentiation error.  Note that we are not making any assumptions on the smoothness of the noise $\eta$, as those are not required for the HD to be applied; in fact, homogeneous differentiators provide estimates of the base signal and its time derivatives with guaranteed accuracy regardless of whether the noise is smooth or non-smooth, as no derivatives of $\eta$ are taken in the dynamics of the HD.

To use the homogeneous differentiator \eqref{eq:DifferentiatorStruct} in applications, it is necessary to appropriately discretise it to allow for a digital implementation that only uses a sampled time-series of the signal $f(t)$. Such a discretisation requires the choice of sampling step size bounds $0<\underline{\theta}\leq \bar{\theta}$. Then, a discrete sequence of sampling times $\left\{ t_s \right\}_{s=1}^{\infty}$ is chosen to satisfy $\underline{\theta}\leq t_{s+1}-t_s = \theta_s \leq \bar{\theta}$, which guarantees $\lim_{s\to \infty}t_s = \infty$.

To be able to obtain provable differentiation guarantees in discrete time, Assumption~\ref{assump:Noise} is replaced for discrete-time homogeneous differentiators by the following:
\begin{assumption}\label{assump:DiscNoise}
The sampled noise  has $n_f+1\in \mathbb{N}$ components, $\eta(t_s) = \sum_{j=0}^{n_f}\eta_j(t_s)$, where $\eta_j(t_s) = \underline{\eta}_j(t_s)+\bar{\eta}_j(t_s),\ j=0,\dots, n_f$. The discretely sampled signals $\underline{\eta}_j(t_s)$ are of the global sampling filtering order $j$ with $j$-th order global integral sampling magnitude $\underline{\epsilon}_j$ (see Definition 3, Section S1.2 in the Supplementary Material); $\bar{\eta}_0=0$ and each noise component $\bar{\eta}_j$ is of the (continuous-time) global filtering order
$j$ with $j$-th order global integral magnitude $\bar{\epsilon}_j$ for $j=0,\dots, n_f$ (see Definition 1, Section S1 in the Supplementary Material). Each noise component ${\bar{\eta}}_j(t)$ is also absolutely continuous with $\left|\dot{\bar{\eta}}_j(t)\right|\leq L_{\eta_j}$ for all $j=0,\dots ,n_f$ for some $L_{\eta_j}>0$.
\end{assumption}
In light of Assumptions~\ref{assump:BaseSignal} and~\ref{assump:DiscNoise}, the discretisation of \eqref{eq:DifferentiatorStruct} has the form of the finite-difference scheme
\begin{equation}\label{eq:DifferentiatorStructDiscrete}
\begin{array}{lll}
& w(t_{s+1}) = w(t_s) + \Omega_{n_d,n_f}\left(w(t_s), z_0(t_s)-f(t_s),L_0,  \lambda \right)\theta_s,\\
&z(t_{s+1}) = z(t_s)+D_{n_d,n_f}\left(w_1(t_s),z(t_s),L_0, \lambda \right)\theta_s+T_{n_d}\left(z(t_s),\theta_s\right),
\end{array}
\end{equation}
where $T_{n_d}\left(z(t_s),\theta_s \right)\in \mathbb{R}^{n_d}$ is a Taylor-like correction term, introduced to preserve the accuracy of the continuous-time scheme after discretisation (see Section S1.2 in the Supplementary Material for details). Due to the presence of $T_{n_d}\left(z(t_s),\theta_s \right)$ on the right-hand side, \eqref{eq:DifferentiatorStructDiscrete} is not a simple Euler discretisation of \eqref{eq:DifferentiatorStruct}.
%\US{For other types of discretisation schemes see \cite{mojallizadeh2021time}.}
Except for $\left\{t_s \right\}_{s=1}^{\infty}$, the required parameters to initialise \eqref{eq:DifferentiatorStructDiscrete} are the same as for the continuous-time scheme \eqref{eq:DifferentiatorStruct}. However, differently from \eqref{eq:DifferentiatorStruct}, the differentiation error bounds for \eqref{eq:DifferentiatorStructDiscrete} depend linearly also on the step-size bound $\bar{\theta}$, which should be chosen to allow for the desired differentiator accuracy, as explained in detail in Section S1 of the Supplementary Material. Similarly to Assumption 
\ref{assump:Noise} for continuous-time differentiators, Assumption~\ref{assump:DiscNoise} enables to obtain rigorous differentiation error guarantees for the discrete-time HD \eqref{eq:DifferentiatorStructDiscrete}.
As the true nature of the noise is unknown, Assumption~\ref{assump:DiscNoise} cannot be verified in practice; nevertheless, as for the continuous-time HD, increasing the filtration degree $n_f$ does not deteriorate the performance, and may even improve the differentiation errors in practice, by filtering out more complex noises (containing more components with higher filtering order). Also, given $n_{d,2} < n_{d,1}$, the $n_{d,1}$-th order differentiator provides a more accurate estimate of the $n_{d,2}$-th derivative than the $n_{d,2}$-th order differentiator \cite{levant2003higher}, provided that the ratio between the noise magnitudes and $L_0$ is less than $1$; this justifies the use of differentiators with a higher order than what is strictly needed.

In this work, we implement the modification of \eqref{eq:DifferentiatorStructDiscrete} proposed in \cite{hanan2021low} and called \emph{discrete $L$-adaptation}, characterised by the additional parameters $q \geq -\frac{1}{n_d+n_f+1}$, which we set to zero ($q=0$) throughout our study, and $k_{\bar{\theta}}>0$ that we choose according to \cite{hanan2021low}; see Section S1.2 of the Supplementary Material for details. This implementation scheme improves the accuracy and significantly lowers chattering (\textit{i.e.}, small-scale high-frequency oscillations) in the output in the absence of noise, while preserving the same differentiation accuracy as the standard HD in the presence of noise.
See \cite{hanan2021low} and Section S1 of the Supplementary Material for additional details.

\subsection{Homogeneous-Differentiator-based attractor reconstruction}
\begin{figure}[h!]
    \centering
    \includegraphics[width=\linewidth]{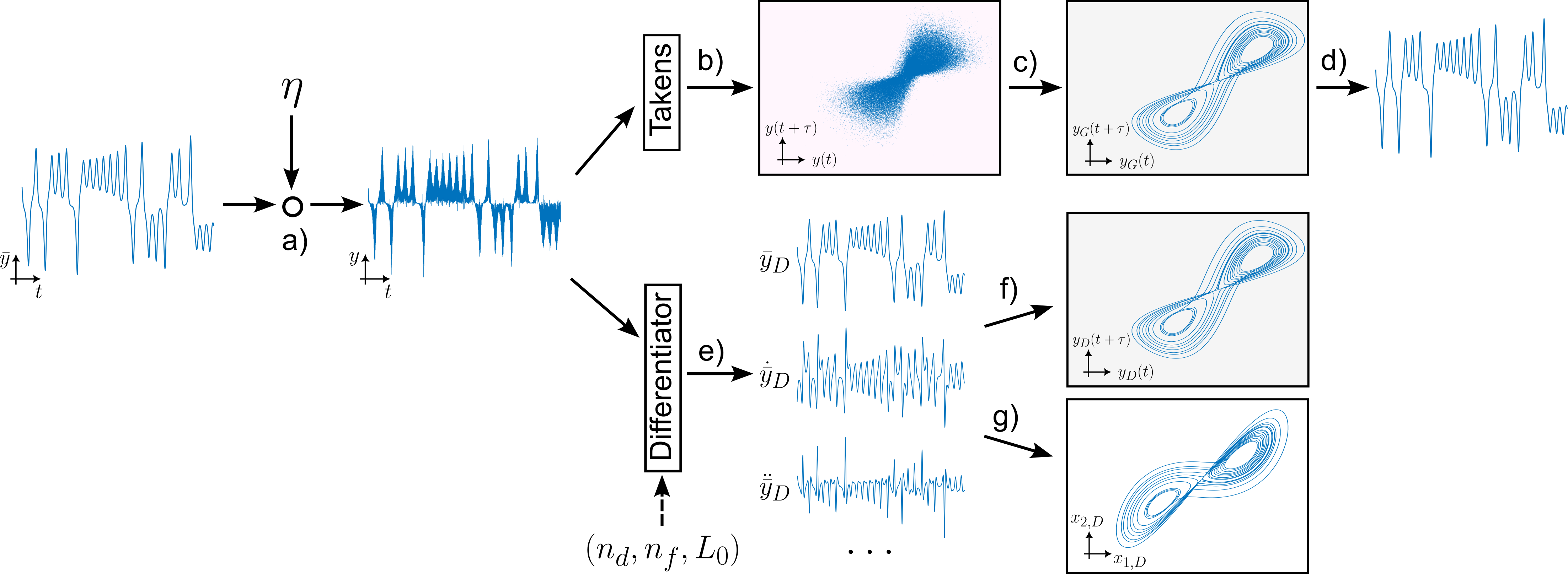}
    \caption{\footnotesize \textbf{Block diagram of the attractor reconstruction procedure relying on signal de-noising achieved with the Homogeneous Differentiator, compared to methods based on time-delay embedding such as Takens' \cite{Takens1981}.} The clean base signal $\bar y(t)$ is corrupted by a noise $\eta(t)$ introduced by the measurement process, which yields the noisy measured signal $y(t)$ (step a). Starting from the noisy signal $y(t)$, methods based on Takens embedding perform attractor reconstruction in the time-delay coordinates $\{y(t), y(t+\tau), \dots, y(t+\ell\tau)\}$ (step b) and subsequent noise reduction on the reconstructed attractor using, \textit{e.g.}, the Schreiber-Grassberger algorithm \cite{SchreiberGrassberger1991,grassberger1993noise} (step c); from the de-noised attractor, one may aim to reconstruct a de-noised estimate of the measured signal \cite{sk2016denoising} (step d). As an alternative methodology, the Homogeneous Differentiator allows us to obtain de-noised estimates of the measured signal and of its derivatives up to the desired order (step e), which immediately enables attractor reconstruction both in the time-delay coordinates (step f) and in the state space (step g). Pink background denotes attractor reconstruction from noisy signals, gray background denotes attractor reconstruction from de-noised signals in the time-delay coordinates, white background denotes attractor reconstruction from de-noised signals in the state space (original coordinates). The output of the Differentiator generated at step e) may be further processed with filters, such as the Savitzky-Golay filter \cite{schafer2011savitzky}, to reduce potential chattering (\textit{i.e.}, small scale oscillations at high frequencies). Signals and attractors in the diagram are generated by the Lorenz system \eqref{eq_lorenz}.}    \label{fig:Differentiator_advantages}
\end{figure}

Consider a dynamical system with state vector $x(t) \in \mathbb{R}^n$, $n \in \mathbb{N}$,
with $t \in \mathbb{R}_+$, and assume that the system has an attractor that we are interested in reconstructing from time-series data.
We assume to have access to measurements of a scalar signal $\bar{y}(t)=\bar{y}(x(t))$ that is a function of the system state $x$; however, the measurement process corrupts the clean base signal $\bar y(t)$ by introducing a noise (or disturbance) $\eta(t)$, so that our actual measurements consist of the noisy scalar signal $y(t)$ (in practice, we typically measure a sampled time-series $y(t_s)$ of the signal $y(t)$ at discrete sampling times $\left\{ t_s \right\}_{s=1}^N$, with $t_s \geq 0$ for all $s$ and $\underline{\theta}\leq t_{s+1}-t_s = \theta_s \leq \bar{\theta}$), where $y$ is some function of both $\bar{y}$ and $\eta$ (Figure~\ref{fig:Differentiator_advantages}a).

Given the noisy measured signal $y(t)$, standard time-delay embedding techniques \cite{Takens1981}, for a specified time delay $\tau$, reconstruct an object in the time-delay coordinates $\{y(t), y(t+\tau), \dots,y(t+\ell \tau)\}$ that is topologically equivalent to the dynamical attractor in the original state-space coordinates (Figure~\ref{fig:Differentiator_advantages}b); $\ell \in \mathbb{N}$ is chosen depending on the expected box-counting dimension of the considered attractor, while the time delay $\tau$ is here determined using custom code, based on the method in \cite{buzug1992optimal}, which computes the value of $\tau$ that maximally separates the trajectories in the time-delay space.
However, direct time-delay embedding from noisy measurements results in a noisy attractor reconstruction, which typically prevents any subsequent topological analysis; hence, a manifold de-noising technique is often used as a post-processing tool to yield a de-noised attractor, which is ideally closer to the true one (Figure~\ref{fig:Differentiator_advantages}c). We consider here the Schreiber-Grassberger algorithm \cite{SchreiberGrassberger1991,grassberger1993noise}, which is the benchmark methodology in the literature; it has been designed to work in the presence of additive white noise with low variance, and no heuristics are explicitly provided for other noise types.
%The Schreiber-Grassberger algorithm needs information about the original dimension of the manifold, which generally requires a much more involved procedure, ---> this applies for any time-delay embedding, you need to choose \ell based on this
%and necessarily the delay $\tau$. ---> this applies for any time-delay embedding

After the attractor de-noising step, additional computations may allow one to retrieve a de-noised estimate of the measured signal (Figure~\ref{fig:Differentiator_advantages}d); for instance, \cite{sk2016denoising} couples a point-wise correction for attractor de-noising using a non-parametric Bayesian approach with convex optimisation.

Our alternative methodology based on the Homogeneous Differentiator has several conceptual and practical advantages, which we systematically study in this work. Applying the HD to noisy time series yields a de-noised estimate of the measurement signal along with all its derivatives up to the desired order (Figure~\ref{fig:Differentiator_advantages}e): this allows us to either reconstruct the attractor using time-delay schemes, by retrieving only the zeroth order derivative and using Takens' embedding \cite{Takens1981} with $\tau$ computed as discussed above (Figure~\ref{fig:Differentiator_advantages}f), or, if the system is observable or functionally observable, retrieve the estimates of higher-order derivatives to reconstruct the attractor in the state-space coordinates (Figure~\ref{fig:Differentiator_advantages}g; see also Figures 
S2 and S9 in the Supplementary Material). Moreover, since the HD provides the de-noised signal (Figure~\ref{fig:Differentiator_advantages}e), no extra steps as in Figure~\ref{fig:Differentiator_advantages}d are needed.

Differently from the the Schreiber-Grassberger algorithm and its derivations, which require multiple computationally expensive operations for each data point,
the HD can be implemented easily and with computational efficiency by a finite-difference scheme.
Also, the HD is effective in the presence of any measurable and essentially bounded noise signal $\eta(t)$, regardless of the specific system generating the base signal $\bar y(t)$; strong theoretical guarantees and error bounds are obtained when the noise is additive, and hence $y(t)=\bar y(t)+\eta(t)$.
Moreover, as it can be seen in Figure~\ref{fig:Differentiator_advantages}, the HD setup only requires to choose three parameters: the number $n_d$ of reconstructed derivatives; a bound $L_0$ on the Lipschitz constant of the $n_d$-th derivative; the filtration degree $n_f$, which can be always increased without deteriorating the accuracy of the estimation \cite{levant2020robust} (see also the discussion in Section~\ref{subsec:hom_filt_diff}). 
An estimate of $L_0$ is typically not available \emph{a priori}; however, to tune the parameter, one can assess the quality of the HD output obtained for increasing values of $L_0$, and note that $L_0$ values that are too large lead to chattering (\textit{i.e.}, small-scale high-frequency oscillations) in the HD output \cite{levant2020robust}; see Supplementary Figure S3.
To further address possible chattering issues \cite{hanan2021low}, the HD output can be post-processed to filter out undesired high-frequency oscillations; to this aim, here we use the Savitzky-Golay (SG) filter \cite{schafer2011savitzky}, which has zero phase, very flat response in the passband, and moderate attenuation in the stopband, and thus leaves the main characteristics of the signal undistorted, while high-frequency components are reduced. Post-processing the HD output with the SG filter appears successful in practice, but the theoretical accuracy guarantees that are valid for the HD output do not extend to the SG filter output (because the SG filter does not enjoy the same homogeneity properties that characterise the HD).

\subsection{Performance metrics}
\label{sec:performance_metrics}

In our numerical experiments based on theoretical models, the model generating $x(t)$ and the clean base ``measurement'' signal $\bar y(t) = \bar y (x(t))$, and hence the true attractor, are known \textit{a priori}, and the 
clean base signal is corrupted by synthetically generated noise $\eta(t)$ of various types, as discussed in Section~\ref{sec:noise_types}. Of course,  no knowledge about the model, the clean signal, the noise or the noise type is used for attractor reconstruction, which is exclusively performed based on noisy discrete samples of the noisy measurement signal $y(t)=y(\bar y(t), \eta(t))$.
However, we use information about the true signal and the true system attractor to conduct comparisons aimed at assessing the quality of the achieved reconstruction.
In particular, in the time-delay coordinates, the reconstruction performance is assessed by comparing the results with those obtained with other attractor reconstruction techniques; in the state space, we can directly compare the results with the true (noise-free) attractor.

We consider a discretely sampled time series $\{y(t_s)\}_{s=1}^N=\{y(k \Delta t)\}_{k=1}^N$, with a constant sampling interval $\Delta t$ (namely, $t_1=\Delta t$ and $t_{s+1}-t_s = \theta_s \equiv \Delta t$ for all $1 \leq s \leq N-1$) such that $\frac{\tau}{\Delta t} \in \mathbb{N}$; the finite total number of measurements is $N\in\mathbb{N}$.
In time-delay coordinates, to quantitatively compare the performance of HD-based attractor reconstruction (Figure~\ref{fig:Differentiator_advantages}f) with the performance of attractor reconstruction from noisy signals followed by Schreiber-Grassberger de-noising (Figure~\ref{fig:Differentiator_advantages}c), we consider the point $\bar{\mathbf{y}}(k)=[\bar{y}(k) \, \, \bar{y}(k+\frac{\tau}{\Delta t}) \,\, \dots \, \, \bar{y}(k+\ell \frac{\tau}{\Delta t})]^\top$ of the system evolution in time-delay coordinates at the discrete time $k\in\mathbb{N}$, $1\leq k \leq N- \ell \frac{\tau}{\Delta t}$ and we compare it with its reconstruction provided through the HD-based method, $\mathbf{y}_D(k)=[y_D(k) \, \, y_D(k+\frac{\tau}{\Delta t})\,\, \dots \, \, y_D(k+\ell \frac{\tau}{\Delta t})]^\top$, and the Schreiber-Grassberger method, $\mathbf{y}_G(k)=[y_G(k) \, \,  y_G(k+\frac{\tau}{\Delta t}) \,\, \dots \, \, y_G(k+\ell \frac{\tau}{\Delta t})]^\top$. The relative reconstruction errors $E_{R,D}$ and $E_{R,G}$, respectively, are given by
\begin{equation}
   E_{R,i}(k) = \frac{\norm{\bar{\mathbf{y}}(k)-\mathbf{y}_i(k)}}{\norm{\bar{\mathbf{y}}(k)}},\quad i\in\{\text{D,G}\},\quad 1\leq k \leq N- \ell \frac{\tau}{\Delta t}\ ,
   \label{eq:E_R_time_delay}
\end{equation}
where $\norm{\cdot}$ is the Euclidean norm.

Similarly, under observability assumptions, to assess the performance of HD-based attractor reconstruction in the state-space coordinates, we consider the relative error
\begin{equation}
   \tilde{E}_R(k) = \frac{\norm{x(k)-x_D(k)}}{\norm{x(k)}}, \quad 1\leq k \leq N- \ell \frac{\tau}{\Delta t} \ .
   \label{eq:E_R_orig_coord}
\end{equation}

In addition, to assess how faithfully some key topological properties of the true attractor are preserved by its reconstruction from noisy measurements, we consider the space-filling properties of the attractor, that is, how densely it fills the whole space; a higher space-filling index is associated with more noise, which lead to more ``blurred'' and ``scattered'' attractors.
For our assessment, we consider the four time series ($\bar{\mathbf{y}}(k)$, to assess the clean base signal; $\mathbf{y}(k)$, to assess the noisy measured signal; $\mathbf{y}_D(k)$, to assess the HD-based reconstruction; and $\mathbf{y}_G(k)$, to assess the Schreiber-Grassberger-de-noised reconstruction) and we employ a methodology based on the box counting approach \cite{wu2020effective}: we partition a region of interest (\textit{e.g.}, the box containing all the points $\bar{\mathbf{y}}(k)=[\bar{y}(k) \, \, \bar{y}(k+\frac{\tau}{\Delta t}) \,\, \dots \, \, \bar{y}(k+\ell \frac{\tau}{\Delta t})]^\top$) into sub-boxes of bin width $b_w$, and we compute the number of sub-boxes that contain at least one point of the considered time series (after a transient, so that the trajectories have converged to the attractor) as a function of the bin width $b_w$, for smaller and smaller values of $b_w$.
The number of non-empty sub-boxes increases when $b_w$ decreases and is upper bounded by the total number $N$ of discrete samples; however, the upper bound is approached faster (and hence the number of non-empty sub-boxes is larger for the same $b_w$) when the space-filling index is higher, meaning that noise-induced corruption is larger.
Further details can be found in Section S2.3 of the Supplementary Material.

\subsection{Considered noise types}
\label{sec:noise_types}
To assess the performance of the HD in reconstructing attractors from noisy measurements $y(t)$, our numerical experiments consider the effect of different types of noises (or disturbances) $\eta(t)$:
\begin{itemize}
\item Gaussian white noises $\eta_{i}(t) \sim \mathcal{N}(0, \sigma_{i}^2)$, independent and identically distributed, with variance $\sigma_1^2 = 0.01$, $\sigma_2^2 = 0.1$, $\sigma_3^2 = 1$ in Section~\ref{sec:Lorenz} and Section~\ref{sec:Epileptor}, and with variance $\sigma_1^2 = 0.001$, $\sigma_2^2 = 0.01$, $\sigma_3^2 = 0.1$ in Section~\ref{sec:Hindmarsh-Rose}; the Schreiber-Grassberger algorithm \cite{SchreiberGrassberger1991,grassberger1993noise} has been developed to address specifically additive Gaussian white noise;
\item harmonic noise of the form
\begin{equation}
    \eta_H(t) = \cos(10000t) - 0.5 \sin(20000t) + 2\cos(70000t),
  \label{eq:harmonic_noise}
\end{equation}
which is not explicitly addressed by the theory of Schreiber-Grassberger-like de-noising algorithms, but has been addressed using the HD in \cite{levant2020robust};
\item unbounded noise of the form
\begin{equation}
    \eta_U(t)= \cos(10000t + 0.7791) + 5\cdot10^{-6}\frac{d^2}{dt^2}[\cos(100t)]^{3/2},    \label{eq:unbounded_noise}
\end{equation}
which is particularly challenging to tackle due to unexpected large deviations, and thus goes way beyond the working assumptions of standard noise reduction methods, but has been tackled via the HD in \cite{levant2020robust}; we consider the same settings as in \cite{levant2020robust} and saturate the noise at magnitude $\pm 100$.
\end{itemize}

All the above noise types can corrupt the measured signal in different ways: we consider the scenario in which noise is additive,
\begin{equation}
    y(t) = \bar{y}(t) + \eta(t), 
\label{eq:additive_noise}
\end{equation}
and the scenario in which noise is multiplicative and state-dependent,
\begin{equation}
    y(t) = [1+\eta(t)]\bar{y}(t). 
\label{eq:multiplicative_noise}
\end{equation} 

Explicit bounds on the estimation error are available within the theory of HDs for additive noise of the form $\eta_H$ and $\eta_U$ \cite{levant2020robust}.
There are currently no theoretical guarantees for the case of multiplicative noise, for which we provide here a thorough numerical investigation.

\section{Reconstructing Attractors of Theoretical Models from Noisy Measurements}
\label{sec:results}

\subsection{Lorenz'63 system: chaotic dynamics}
\label{sec:Lorenz}

A widely adopted benchmark for attractor reconstruction is the Lorenz'63 system, given by the ordinary differential equations \cite{Lorenz63}
\begin{equation}\label{eq_lorenz}
    \begin{cases}
        \dot{x}_1(t) = \sigma[x_2(t) - x_1(t)]\\
        \dot{x}_2(t) = r x_1(t) - x_2(t) - x_1(t) x_3(t)\\
        \dot{x}_3(t) = x_1(t) x_2(t) - b x_3(t)
    \end{cases}    
\end{equation}
with state vector $x(t)=[x_1(t)\,\,x_2(t)\,\,x_3(t)]^\top$. Choosing the parameter values as $\sigma=10$, $b=8/3$ and $r=28$ yields a chaotic attractor (associated with a deterministic non-periodic flow) \cite{Lorenz63}. From noisy measurements $y(t)$ of the clean base signal $\bar y(t)=\bar y(x(t))=x_1(t)$, we wish to reconstruct the projection of the system's chaotic attractor in the $\{x_1, x_2\}$ subspace (which from now we call the attractor); this is always possible because the system given by \eqref{eq_lorenz} and measurements of $x_1$ is functionally observable everywhere with respect to the functional $\zeta(x)=x_2$, as discussed in \cite[Section IV.A]{montanari2022functional}.

\begin{figure}[ht!]
    \centering
    \includegraphics[width=0.6\linewidth]{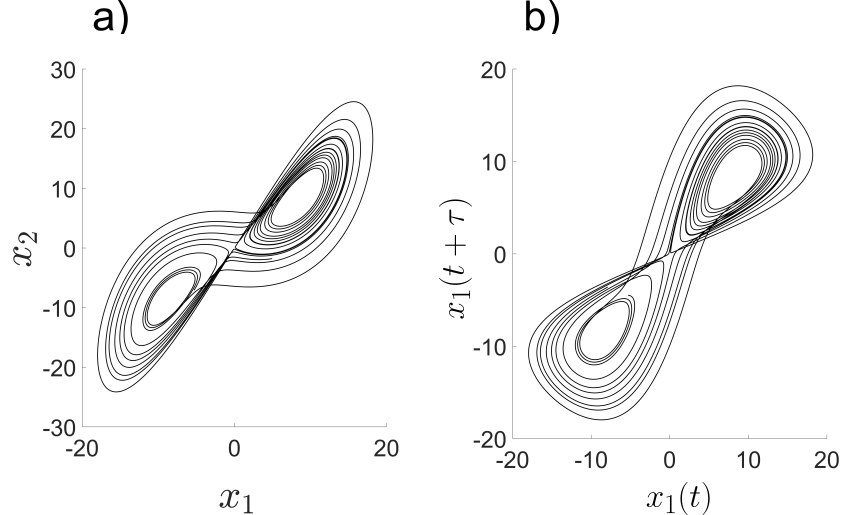}
    \caption{\footnotesize \textbf{The true attractor of the Lorenz system in the state-space coordinates vs. the topologically equivalent attractor in the time-delay coordinates.} True chaotic attractor of the Lorenz system~\eqref{eq_lorenz} shown a) in the two-dimensional subspace $\{x_1, x_2\}$ and b) in the time-delay coordinates $\{x_1(t), x_1(t+\tau) \}$, with $\tau = 0.1$.}
    \label{fig:Lorenz_dynamics}
\end{figure}

As our ``ground truth'' for comparison, the true system attractor is shown in the $\{x_1, x_2\}$ subspace in Figure~\ref{fig:Lorenz_dynamics}a, and in the time-delay coordinates $\{x_1(t), x_1(t+\tau)\}$, with $\tau=0.1$, in Figure~\ref{fig:Lorenz_dynamics}b.

\begin{figure}[h!]
    \centering
    \includegraphics[width=0.95\linewidth]{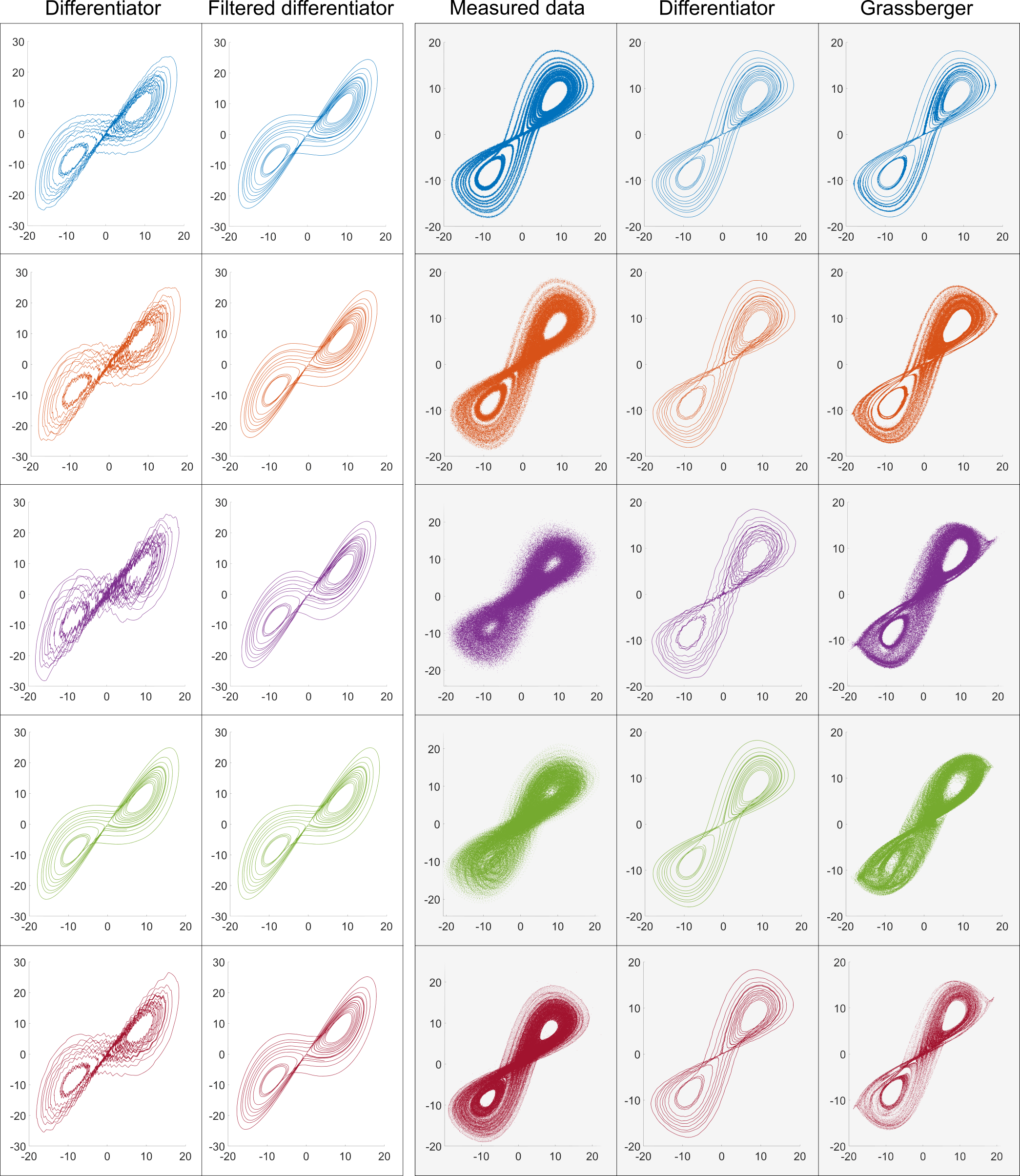}
    \caption{\footnotesize \textbf{The Differentiator enables improved de-noising and faithful attractor reconstruction for the Lorenz system affected by various types of additive noise, both in the state-space and in the time-delay coordinates.} Reconstructed attractor of the Lorenz system \eqref{eq_lorenz} from measurements of $x_1$ affected by \textit{additive} noise. The Differentiator parameters are set as $n_d=2$, $n_f=3$, $L_0=3.75\cdot 10^4$. \textbf{White, left-most columns:} reconstruction of the attractor in the subspace $\{x_1(t), x_2(t) \}$, using the de-noised signal provided directly by the Differentiator (first column) and the Differentiator output post-filtered with the Savitzky-Golay filter (second column). \textbf{Grey, right-most columns:} reconstruction of the attractor in the time-delay coordinates $\{ x_1(t), x_{1}(t+\tau) \}$  using directly the noisy measured data (third column), the de-noised signal provided by the Differentiator (fourth column), and the noisy measured data followed by manifold de-noising via the Schreiber-Grassberger method (fifth column). \textbf{Rows:} noise types.
    From top to bottom, Gaussian noise $\eta_1(t) \sim \mathcal{N}(0, 0.01)$ (blue), $\eta_2(t) \sim \mathcal{N}(0, 0.1)$ (orange), $\eta_3(t) \sim \mathcal{N}(0, 1)$ (violet); harmonic noise $\eta_H(t)$ as in \eqref{eq:harmonic_noise} (green); unbounded noise $\eta_U(t)$ as in \eqref{eq:unbounded_noise} (red). }
    \label{fig:Lorenz additive noise}
\end{figure}

Figure~\ref{fig:Lorenz additive noise} compares the reconstructed attractors obtained with different methodologies, when measurements are affected by additive noise $\eta(t)$ as in \eqref{eq:additive_noise} considering, from top to bottom, Gaussian noise $\eta_1(t) \sim \mathcal{N}(0, 0.01)$, $\eta_2(t) \sim \mathcal{N}(0, 0.1)$, $\eta_3(t) \sim \mathcal{N}(0, 1)$, harmonic noise $\eta_H(t)$ as in \eqref{eq:harmonic_noise}, and unbounded noise $\eta_U(t)$ as in \eqref{eq:unbounded_noise}; the resulting noisy time series are reported in the Supplementary Figure S1.

Given the measured signal $y(t)$, the HD-based method allows us to immediately reconstruct the attractor in the $\{x_1, x_2 \}$ subspace (Figure~\ref{fig:Lorenz additive noise}, left-most set of columns), because it provides de-noised estimates of $y$ (which is an approximation of $x_1$) and of $\dot y$ (which is an approximation of $\dot x_1$) and then, from the first equation in system \eqref{eq_lorenz}, we can approximate $x_2=\frac{\dot{x}_1}{\sigma}+x_1$ as $\frac{\dot{y}}{\sigma}+y$ (see details in the Supplementary Figure S2).
The first column of Figure~\ref{fig:Lorenz additive noise} shows the reconstructed attractor in the $\{x_1, x_2 \}$ subspace, after processing the measured signal $y(t)$ with the Differentiator. With Gaussian noise having variance $\sigma_1^2=0.01$ and with harmonic noise, the faithfulness to the true attractor (Figure~\ref{fig:Lorenz_dynamics}a) is already striking. With the other noise types, chattering slightly alters the shape of the reconstructed attractor: in these cases, as shown in the second column, post-processing with a simple low-pass SG filter drastically improves the result, yielding an attractor that is almost identical to the true one.
In the time-delay coordinates, we can compare the reconstruction obtained directly from the noisy measured signal (direct embedding) with the HD-based reconstruction obtained by de-noising the signal with the Differentiator (without post-filtering with the SG filter) and with the reconstruction obtained by applying, after direct embedding, the Schreiber-Grassberger de-noising method (Figure~\ref{fig:Lorenz additive noise}, right-most set of columns).
The attractor reconstructed using the Differentiator, even without SG filtering, is already very close to the true attractor in the time-delay coordinates (Figure~\ref{fig:Lorenz_dynamics}b) and drastically improves over the results with the other two methods. In fact, performing direct time-delay embedding of measured data without noise reduction, as often done, \textit{e.g.}, in tools for topological analysis \cite{tauzin2020giottotda}, may preserve some topological properties of the attractor (such as the number of holes, which are still distinguishable in most of the considered cases), but the points are scattered due to noise and the resulting attractor is blurry. While alleviating this phenomenon, the Schreiber-Grassberger method induces deformations in some locations (such as the little ``horns'' that are clearly visible in case of intense Gaussian noise or unbounded noise); in any case, the achieved noise suppression is not as neat as that provided by the Differentiator.

\begin{figure}[h!]
    \centering
    \includegraphics[width=0.95\linewidth]{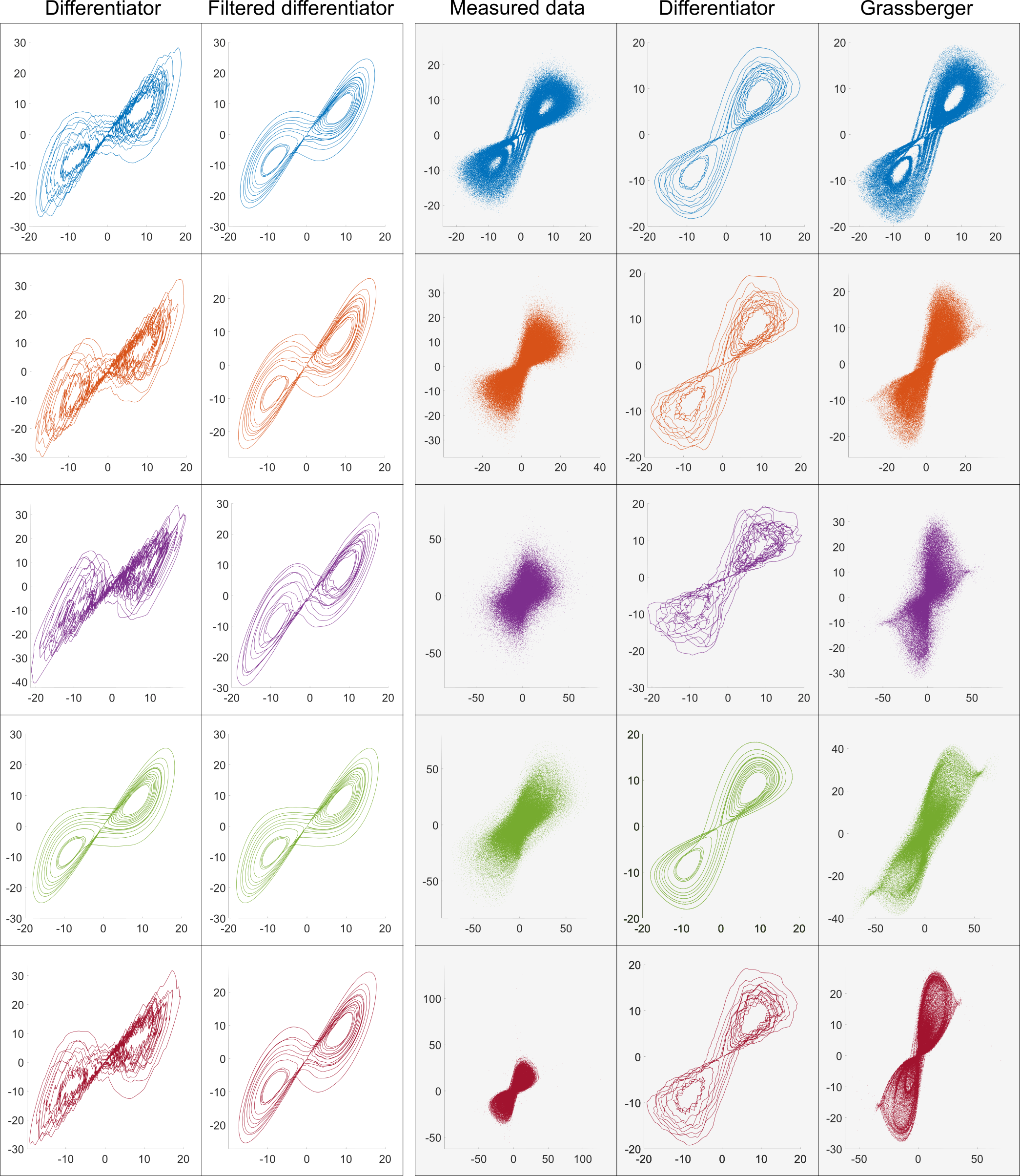}
    \caption{\footnotesize
    \textbf{The Differentiator enables improved de-noising and faithful attractor reconstruction for the Lorenz system affected by various types of multiplicative noise, both in the state-space and in the time-delay coordinates.} Reconstructed attractor of the Lorenz system \eqref{eq_lorenz} from measurements of $x_1$ affected by \textit{multiplicative} noise. The Differentiator parameters are set as $n_d=2$, $n_f=3$, $L_0=3.75\cdot 10^4$. \textbf{White background:} attractor reconstruction in the subspace $\{x_1(t), x_2(t) \}$, using the Differentiator output (first column) and the Differentiator output post-filtered with the Savitzky-Golay filter (second column). \textbf{Grey background:} attractor reconstruction in the time-delay coordinates $\{ x_1(t), x_{1}(t+\tau) \}$  using the noisy measured data (third column), the Differentiator output (fourth column), and the noisy measured data followed by manifold de-noising via the Schreiber-Grassberger method (fifth column). \textbf{Rows:} noise types.
    From top to bottom, Gaussian noise $\eta_1(t) \sim \mathcal{N}(0, 0.01)$ (blue), $\eta_2(t) \sim \mathcal{N}(0, 0.1)$ (orange), $\eta_3(t) \sim \mathcal{N}(0, 1)$ (violet); harmonic noise $\eta_H(t)$ as in \eqref{eq:harmonic_noise} (green); unbounded noise $\eta_U(t)$ as in \eqref{eq:unbounded_noise} (red); for measured data, the plot is significantly rescaled in the case of unbounded noise to include some scattered points that are very far from the bulk.}
    \label{fig:Lorenz multiplicative noise}
\end{figure}

Figure~\ref{fig:Lorenz multiplicative noise} compares the reconstructed attractors obtained with different methodologies, when measurements are subject to multiplicative noise $\eta(t)$ as in \eqref{eq:multiplicative_noise}. Even in the absence of theoretical guarantees, the Differentiator reconstructs the attractor in the $\{x_1, x_2 \}$ subspace remarkably well, especially if post-filtered (Figure~\ref{fig:Lorenz multiplicative noise}, left-most set of columns).
When performing time-delay embedding, the Differentiator (even without SG post-filtering) is the only method that always retains a clear topological structure of the attractor; except for the case with the mildest Gaussian noise, both reconstructions via direct embedding and via the Schreiber-Grassberger noise-reduction technique lead to a strong deformation of the attractor and lose topological information (\textit{e.g.}, about holes). Whenever the Differentiator results are subject to chattering, such as in the case of intense Gaussian noise or unbounded noise, post-filtering with the SG filter would further improve the smoothness of the reconstruction.

For a quantitative comparison between the different methods, we compute the relative errors $E_{R,i}$ as in \eqref{eq:E_R_time_delay}.
Table~\ref{tab:performance_lorenz}, top, summarises the most relevant statistics of the distribution of $E_{R,i}(k)$ (for the full distributions of $E_{R,i}(k)$, we refer to the Supplementary Figure S5). The Differentiator significantly outperforms Schreiber-Grassberger noise reduction, which already suppresses the discrepancies associated with direct embedding of noisy measurements, by providing more than tenfold error reduction in the case of Gaussian and unbounded noise and an even stronger reduction in the case of harmonic noise.
The performance of the Differentiator further improves after SG post-filtering. The advantage of the Differentiator is even more pronounced in the case of multiplicative noise. 

\begin{figure}[h!]
    \centering
    \includegraphics[width=0.95\linewidth]{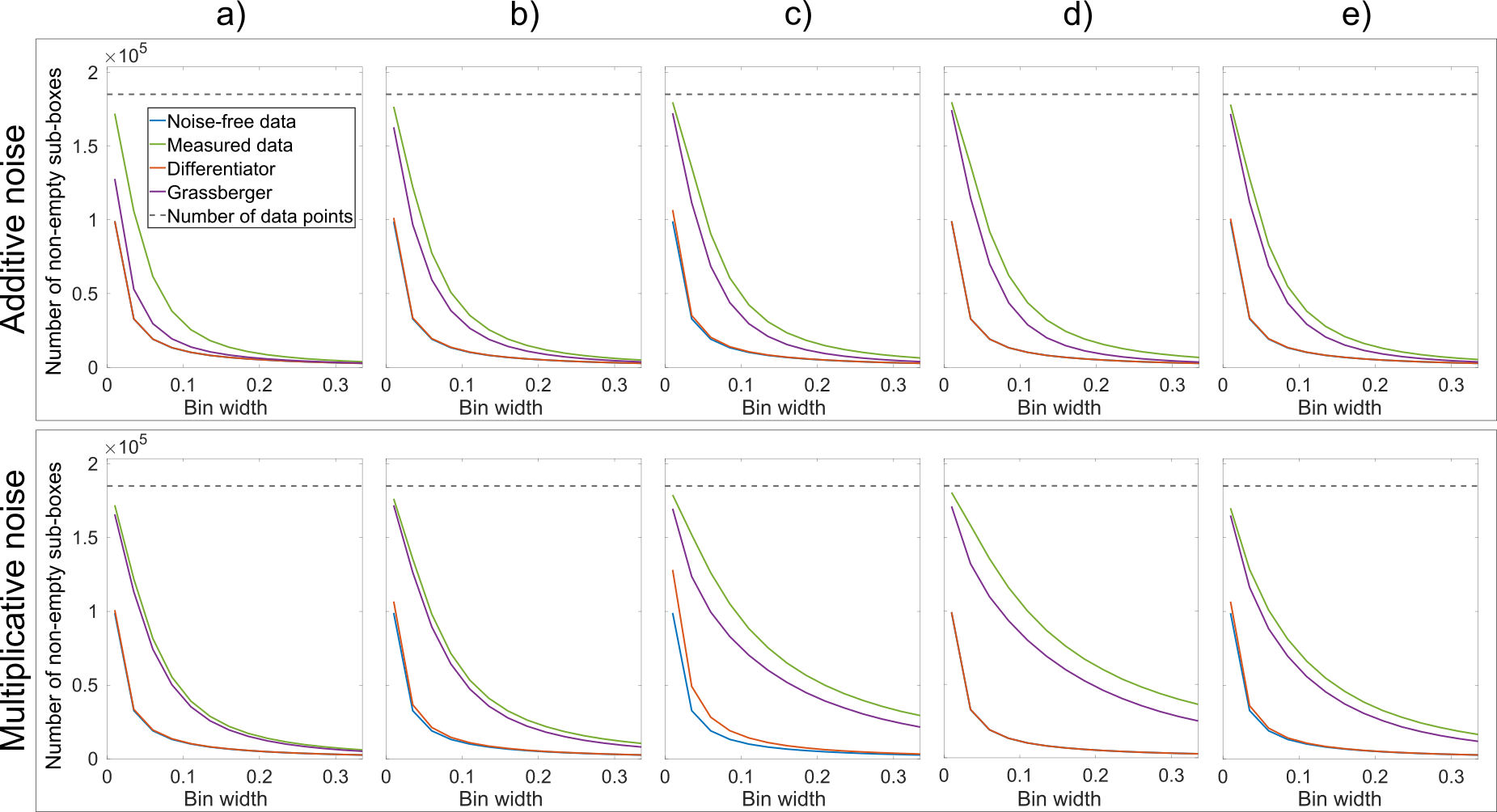}
    \caption{\footnotesize \textbf{For the Lorenz system, quantification through the space-filling index of the reconstruction quality, which is higher if the index value is closer to that obtained with noise-free data. In all simulated cases, the reconstruction based on the Differentiator outperforms the alternative approaches, both when the noise is additive and (in an even more pronounced way) when the noise is multiplicative.} Space-filling index (\textit{i.e.}, number of non-empty sub-boxes) as a function of the bin width, computed as described in Section~\ref{sec:performance_metrics}, for the time-delay embedding of the reconstructed attractor of the Lorenz system \eqref{eq_lorenz} based on the noise-free signal data (blue), the noisy measured signal data (green), the HD-based reconstruction (red) and the noisy reconstruction after Schreiber-Grassberger de-noising (violet). The blue and red line often overlap, which attests to the high quality of the HD-based attractor reconstruction; a larger space-filling index for a given bin width denotes poorer quality of the reconstruction. The horizontal dashed line denotes the total number $N$ of data points, which is the theoretical maximum number of non-empty sub-boxes. We consider different types of noise corrupting the measured signal: \textbf{a)} Gaussian white noise $\eta_1$; \textbf{b)} Gaussian white noise $\eta_2$; \textbf{c)} Gaussian white noise $\eta_3$; \textbf{d)} harmonic noise $\eta_H$; \textbf{e)} unbounded noise $\eta_U$. \textbf{Top row:} additive noise. \textbf{Bottom row:} multiplicative noise.}
    \label{fig:Lorenz space-filling}
\end{figure}

HD-based attractor reconstruction is also extremely robust in preserving the attractor's topological properties in the face of noise. Figure~\ref{fig:Lorenz space-filling} captures the space-filling properties of the reconstructed attractors, as described in Section~\ref{sec:performance_metrics}. For all noise types, the Differentiator preserves the pattern of the true attractor: the red line, associated with the HD-based reconstruction, is almost perfectly overlapping with the blue line, associated with the noise-free signal. The space-filling index obtained with direct embedding (measured data without noise reduction) and with Schreiber-Grassberger noise reduction is much larger: this indicates scattering and loss of the space-filling properties of the attractor, which may impair a subsequent analysis based on topological properties. HD-based reconstruction preserves not only the global space-filling properties of the whole attractor, but also the persistence of specific topological elements, such as holes, as shown in the Supplementary Figures S7 and S8.

Finally, the Differentiator has an impressive computational efficiency: as shown in Table~\ref{tab:performance_lorenz}, bottom, on a total of 200,001 processed data points, it performs both de-noising and attractor reconstruction in less than $1$ second for all the considered cases (and at most $1$ second when also including SG post-processing), while the Schreiber-Grassberger method always requires several hours (at least $7$).

\begin{table}[h!]
\centering \footnotesize
\begin{tabular}{ll|lllll|lllll|}
\multicolumn{1}{l}{} &                         & \multicolumn{5}{c|}{Additive noise}                   & \multicolumn{5}{c|}{Multiplicative noise}             \\ \cline{3-12} 
         $E_R$           &    Method   & $\eta_1$ & $\eta_2$ & $\eta_3$ & $\eta_H$ & $\eta_U$ & $\eta_1$ & $\eta_2$ & $\eta_3$ & $\eta_H$ & $\eta_U$ \\ \hline
\multirow{3}{*}{Mean} & Diff.  & 0.0075  & 0.020  & 0.046   & 0.00053 & 0.014 & 0.014 & 0.037 & 0.097 & 0.0011 & 0.030   \\
                     & Diff. SG   & 0.0053 & 0.012  & 0.027  & 0.00045 & 0.0048 & 0.0059 & 0.016 & 0.039 & 0.001 & 0.0092 \\
                     & Grassb.  & 0.022  & 0.086  & 0.26 & 0.52  & 0.22 & 0.091 & 0.28 & 0.81 & 1.3 & 0.75 \\ \hline
\multirow{3}{*}{Median} & Diff. & 0.0026 & 0.0068   & 0.018   & 0.00035 & 0.0053  & 0.012 & 0.032 & 0.086 & 0.00081 &  0.026  \\
                     & Diff. SG. & 0.0035 & 0.0082  & 0.016   & 0.00033 & 0.0034 & 0.005 & 0.013 & 0.031 & 0.00074 & 0.0065  \\
                     & Grassb.  & 0.0078 & 0.032 & 0.099 & 0.19 & 0.094 & 0.063 & 0.21 & 0.65 & 1.1 & 0.71  \\ \hline
\multirow{3}{*}{Max.} & Diff. & 0.36  & 0.80   & 1.3  & 0.014 & 0.6 & 0.068 & 0.22 & 0.43 & 0.0082 & 0.14     \\
                     & Diff. SG. & 0.045  & 0.24   & 0.48    & 0.0051  & 0.051 & 0.035 & 0.13 & 0.35 & 0.0079  & 0.094 \\
                     & Grassb.  & 2.5 & 6.5 & 24  & 33  & 12  & 3.3 & 25 & 66 & 17 & 26 \\ \hline \hline
\multirow{3}{*}{$T_\text{comp}$} 
& Diff. \quad \; $[\mathbf{s}]$   & 0.92 &  0.84 & 0.83 & 0.87 & 0.86 & 0.90 & 0.84 & 0.86 & 0.91 & 0.83 \\
& Diff. SG  $[\mathbf{s}]$ & 1.0 & 0.97 & 1.0 & 0.90 & 0.96 & 0.99 & 0.97 & 1.0 & 0.94 & 0.93 \\
& Grassb. \ $[\mathbf{h}]$ & 7.33   & 8.75  & 8.22 & 11.75  & 17.12 & 8.80 & 9.40 & 9.80 & 11.2 & 8.91 \\ \hline
\end{tabular}
\caption{\textbf{Lorenz system: Reconstruction error and computational time achieved in numerical experiments show the efficiency and faithfulness of Differentiator-based attractor reconstruction.} \textbf{Top:} Summary statistics of relative errors $E_R$ in \eqref{eq:E_R_time_delay} for attractor reconstruction from measurements subject to different noise types, using the Differentiator, the Differentiator with SG post-filtering, and the Schreiber-Grassberger noise reduction method, for the Lorenz system \eqref{eq_lorenz}. \textbf{Bottom:} Computational time $T_\text{comp}$ for attractor reconstruction (on a Dell Inspiron 16 laptop with 16 GB RAM and 1.90 GHz Intel i5-1340P core processor running Windows), reported in \textbf{seconds} $[s]$ for the Differentiator and in \textbf{hours} $[h]$ for the Schreiber-Grassberger algorithm.}
\label{tab:performance_lorenz}
\end{table}

\subsection{Hindmarsh-Rose model: multiple dynamical regimes}
\label{sec:Hindmarsh-Rose}
We consider the phenomenological HR model of neural dynamics proposed by Hindmarsh and Rose in \cite{hindmarsh1984model}.
The adimensionalised equations of the model, resulting from a simplification of the biophysical Hodgkin-Huxley neuronal model \cite{HH1952} and a generalisation of the FitzHugh-Nagumo single-neuron model \cite{izhikevich2007dynamical}, are
\begin{equation}\label{eq:HR}
    \begin{cases}
        \dot x_1(t) = x_2(t) - a x_1^3(t) + b x_1^2(t) - x_3(t) + I\\
        \dot x_2(t) = c - d x_1^2(t) - x_2(t)\\
        \dot x_3(t) =  r[s(x_1(t) - x_R) - x_3(t)]
    \end{cases}
\end{equation}
where $x_1$ represents the membrane potential, $x_2$ the fast recovery current, and $x_3$ the slow adaptation current. The parameter $I$ represents an externally injected current, either due to stimulation during an experiment or during synaptic activity \textit{in vivo}; the other parameters relate to geometric properties of the dynamics and allow to switch between various dynamic regimes (i.e., different long-term behaviours of the system solutions) \cite{hindmarsh1984model}.
In fact, depending on the parameter values, the model can exhibit a wide range of dynamical behaviours \cite{storace2008hindmarsh, sutulovic2024efficient}; we set $a = 1$, $c = 1$, $d = 5$, $s = 4$, $x_R = -8/5$, $r=0.01$ and we choose $I$ and $b$ so as to obtain five different regimes: quiescence ($I=2.2$, $b=3.2$), tonic spiking ($I=2.5$, $b=3$), square-wave bursting ($I=3$, $b=2.7$), plateau bursting ($I=4$, $b=2.5$) and chaotic bursting ($I=2.7$, $b=3.03$). The distinctive time evolution of the state variable $x_1$ corresponding to the five dynamic regimes is illustrated in the first column of Figure~\ref{fig:HR additive noises}; see also Section S3 of the Supplementary Material.

\begin{figure}[ht!]
    \centering
    \includegraphics[width=0.9\linewidth]{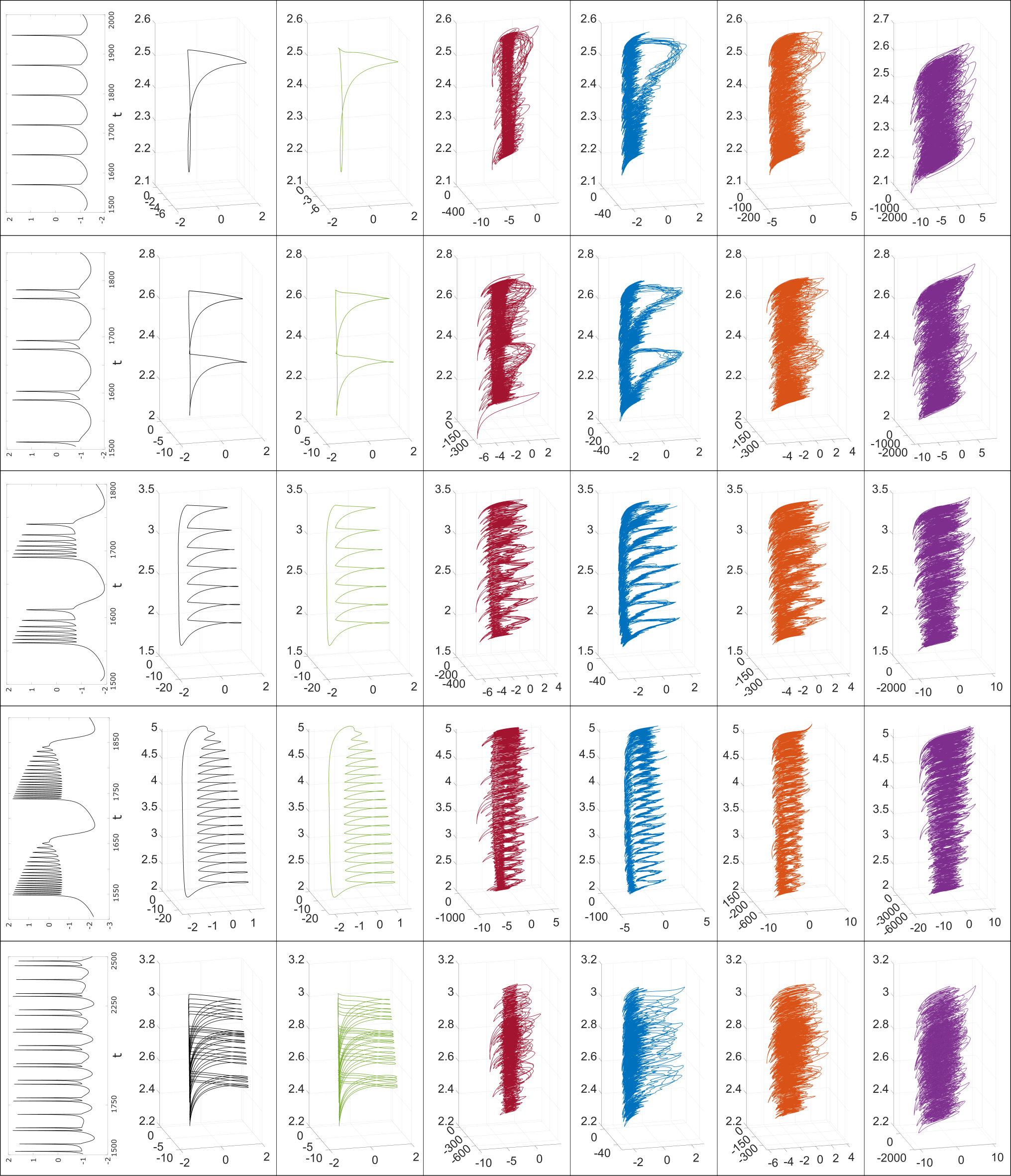}
    \caption{\footnotesize  \textbf{The Differentiator enables de-noising and faithful attractor reconstruction in different regimes for the HR system affected by various types of multiplicative noise, in the state-space coordinates.} Reconstructed attractor of the HR model \eqref{eq:HR} from measurements of $x_3$ affected by \textit{multiplicative} noise obtained through the HD-based method. The Differentiator parameters are set as $n_d=3$, $n_f=9$, $L_0=10$. \textbf{Rows, from top to bottom:} quiescent, tonic spiking, square-wave bursting, plateau bursting and chaotic bursting regimes. \textbf{Columns, from left to right:} distinctive time evolution of the system state variable $x_1$ for the considered regime; true attractor of the system in the $\{x_1, x_2, x_3\}$ space (black); reconstruction of the attractor (in the same space) from measurements corrupted by multiplicative noise $\eta_H(t)$ as in \eqref{eq:harmonic_noise} (green); $\eta_U(t)$ as in \eqref{eq:unbounded_noise} (red); $\eta_1(t) \sim \mathcal{N}(0, 0.001)$ (blue); $\eta_2(t) \sim \mathcal{N}(0, 0.01)$ (orange); $\eta_3(t) \sim \mathcal{N}(0, 0.1)$ (violet).}
    \label{fig:HR multiplicative noises}
\end{figure}

The flexibility of the HR model allows us to test our proposed HD-based attractor reconstruction methodology on a wide variety of dynamic regimes.
Moreover, the HR model poses a challenge, because its dynamics evolve on multiple time scales, and therefore employing a simple low-pass filter to de-noise the output signal may lead to discarding important components of it. Suppose that we do not have direct access to measurements of the membrane potential $x_1$, for instance because the neural structure is so thin that inserting electrodes is extremely challenging, or because we are conducting a study in vivo and patch-clamp recording \cite{hamill1981improved} is too invasive. We assume, however, to be able to measure the state $x_3$, a slow current variable that physiologically describes the influx of $\text{Ca}^{2+}$ \cite{gu2013biological} and can be quantified via imaging \cite{grienberger2012imaging}; hence, the system is locally observable everywhere \cite[Section IV.C]{montanari2022functional} and, consequently, the estimates of subsequent derivatives provided by the Differentiator can be used to invert the dynamics (see Supplementary Figure S9 for details) and reconstruct the attractor in the whole $\{x_1,x_2,x_3\}$ state space.
We no longer conduct a comparison with direct embedding, which would not be intelligible due to the noise, nor with the method relying on the Schreiber-Grassberger de-noising algorithm, which would be computationally prohibitive; we just employ the HD-based method to directly reconstruct the system attractor in the state-space coordinates, for the five different regimes. In our numerical experiments, we consider the Gaussian white noises $\eta_{i}(t) \sim \mathcal{N}(0, \sigma_i^2)$, $i=1,2,3$, independent and identically distributed, with variances $\sigma_1^2 = 0.001$, $\sigma_2^2=0.01$, $\sigma_3^2=0.1$, as well as the harmonic noise $\eta_H(t)$ and the unbounded noise $\eta_U(t)$. 
Figure~\ref{fig:HR additive noises} shows the results with additive noise and Figure~\ref{fig:HR multiplicative noises} with multiplicative noise. Except for the cases of additive and multiplicative Gaussian white noise with large variance and of multiplicative unbounded noise, the attractor is faithfully reconstructed for all the regimes, including chaotic bursting; in particular, the number of lobes of the attractor is preserved.

For a quantitative performance assessment, the distributions and the summary statistics of relative errors for each dynamical regime are provided in the Supplementary Figures S20 and S21. The relative errors $\tilde{E}_R$ in the state space coordinates, as per \eqref{eq:E_R_orig_coord}, are one or two orders of magnitude higher than the errors $E_R$ in the time-delay coordinates as per \eqref{eq:E_R_time_delay} obtained for the Lorenz system (Table~\ref{tab:performance_lorenz}).
This is expected, since the attractor reconstruction in the original state-space coordinates requires the estimation of higher-order derivatives, which is intrinsically more challenging (the problem is fundamentally ill-posed, as we discussed in the Introduction); also the theoretical error bounds in the case of additive noise are larger for higher-order derivatives than for zeroth-order derivatives (see Theorem 3 in the Supplementary Material). Nonetheless, as it can be seen in Figures~\ref{fig:HR additive noises} and \ref{fig:HR multiplicative noises}, the attractor reconstruction enables us to clearly distinguish between the dynamical regimes (\textit{e.g.}, distinguish spiking from bursting).

A single attractor reconstruction for the HR model \eqref{eq:HR} with the HD-based method takes less than two minutes for 10,000,001 time points, while it would be computationally prohibitive with the Schreiber-Grassberger algorithm, which takes more than 7 hours for 200,001 time points (\textit{cf.} Table~\ref{tab:performance_lorenz}) and, since one needs to apply a “correction" to each of these points in the appropriate time-delay coordinates, scales at least linearly with the number of data points \cite[Section V.B]{grassberger1993noise}. 

\begin{figure}[ht!]
    \centering
    \includegraphics[width=0.9\linewidth]{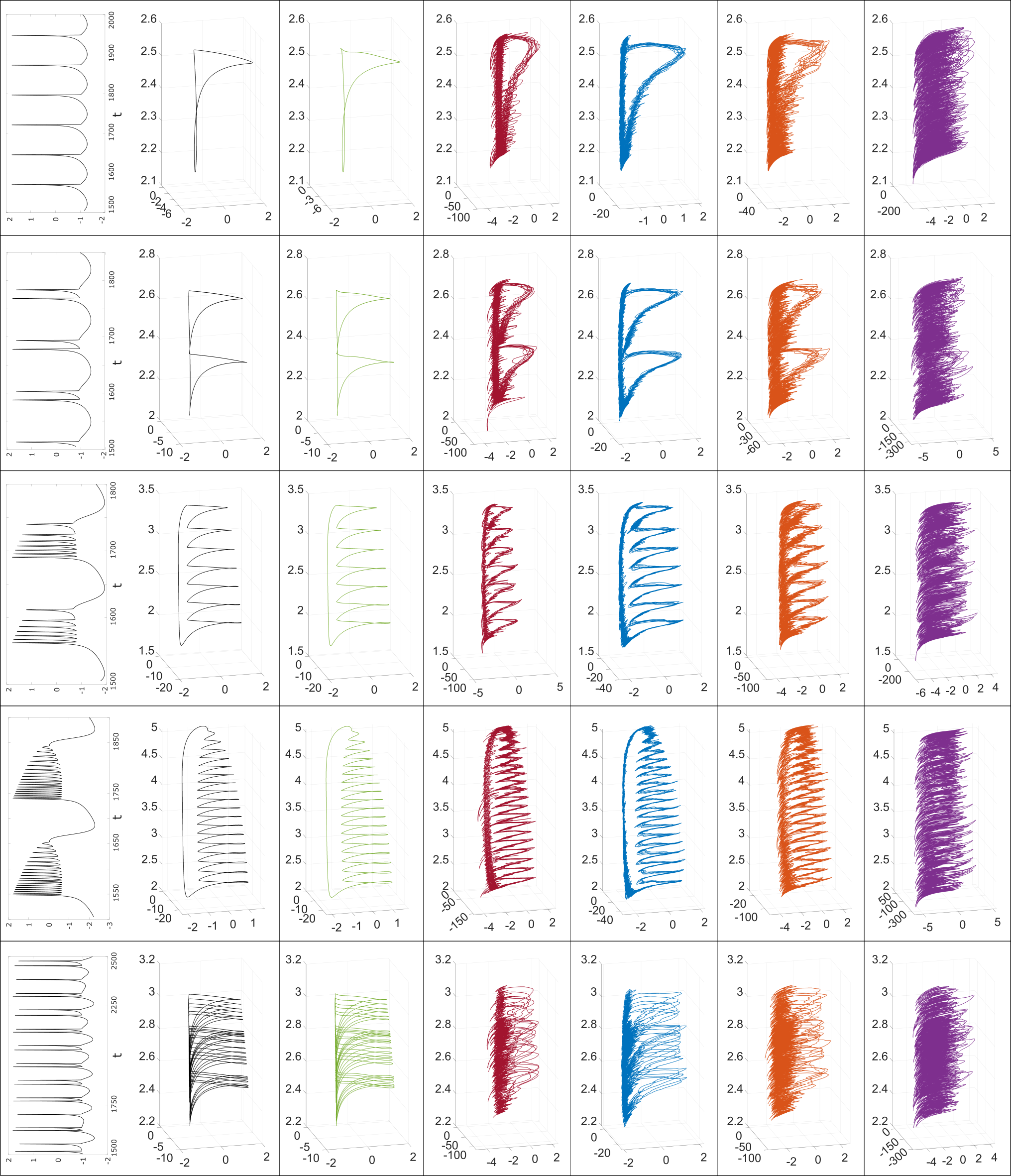}
    \caption{\footnotesize \textbf{The Differentiator enables de-noising and faithful attractor reconstruction in different regimes for the HR system affected by various types of additive noise, in the state-space coordinates.} Reconstructed attractor of the HR model \eqref{eq:HR} from measurements of $x_3$ affected by \textit{additive} noise obtained through the HD-based method. The Differentiator parameters are set as $n_d=3$, $n_f=9$, $L_0=10$. \textbf{Rows, from top to bottom:} quiescent, tonic spiking, square-wave bursting, plateau bursting and chaotic bursting regimes. \textbf{Columns, from left to right:} distinctive time evolution of the system state variable $x_1$ for the considered regime; true attractor of the system in the $\{x_1, x_2, x_3\}$ space (black); reconstruction of the attractor (in the same space) from measurements corrupted by additive noise $\eta_H(t)$ as in \eqref{eq:harmonic_noise} (green); $\eta_U(t)$ as in \eqref{eq:unbounded_noise} (red); $\eta_1(t) \sim \mathcal{N}(0, 0.001)$ (blue); $\eta_2(t) \sim \mathcal{N}(0, 0.01)$ (orange); $\eta_3(t) \sim \mathcal{N}(0, 0.1)$ (violet).}
    \label{fig:HR additive noises}
\end{figure}

\subsection{Epileptor model: measuring multiple variables to monitor epileptic seizures}
\label{sec:Epileptor}
We consider a large-scale phenomenological model representing seizure dynamics in epilepsy: the Epileptor \cite{jirsa2014nature, el2020epileptor}, a mean-field model that builds upon the HR model \eqref{eq:HR} and aims at reproducing the mean effect of an action potential on a block of neurons. The model equations are
\begin{equation}
\begin{cases}
\dot x_1(t) = x_2(t) - f_1(x_1(t),x_4(t)) - x_3(t) + I_{1}\\
\dot x_2(t) = r_2 - 5x_1^2(t) - x_2(t)\\
\dot x_3(t) = \frac{1}{\tau_0} [4(x_1(t) - r_1) - x_3(t)]\\
\dot x_4(t) = -x_5(t) + x_4(t) - x_4^3(t) + I_{2} + 2u(t) - 0.3(x_3(t) - 3.5)\\
\dot x_5(t) = \frac{1}{\tau_2} [-x_5(t) + f_2(x_4(t))] \\
\dot{u}(t) = -\gamma [u(t)-0.1 x_1(t)]
\end{cases}
\label{eq:Epileptor}
\end{equation}
with time scale constants $\tau_0=2857$, $\tau_2=10$ and $\gamma=0.01$, system parameters $r_{1}=-1.6$, $r_{2}=1$, $I_{1}=3.1$ and $I_{2} = 0.42$, and coupling functions defined as
\begin{equation}
    \begin{aligned}
        f_1(x_1(t),x_4(t)) &= 
        \begin{cases}
        x_1^3(t) - 3x_1^2(t), \quad & x_1(t) < 0, \\
        \{x_4(t) - 0.6[x_3(t) - 4]^2\}x_1(t), \quad & x_1(t) \geq 0, \\
        \end{cases} \\
        f_2(x_4(t)) &= 
        \begin{cases}
        0, \quad & x_4(t) < -0.25, \\
        6[x_4(t) + 0.25], \quad & x_4(t) \geq -0.25. \\
        \end{cases}
    \end{aligned}
\end{equation}

The state variables $x_i$, $i=1 \dots 5$, describe the evolution of three subsystems evolving on different time scales: $x_1$ and $x_2$ govern the system's oscillatory behaviour, $x_4$ and $x_5$ lead to the spikes and wave components that characterise seizure-like events, while $x_3$ represents a slow permittivity variable that drives the system close to the seizure threshold. Finally, $u$ is a low-pass filter variable.

With the chosen parameter values, the Epileptor model generates seizure-like events, as the time evolution of $x_1(t)+x_4(t)$ bears resemblance to experimentally recorded field potentials \cite{jirsa2014nature}; see, \textit{e.g.}, the signal $\bar y = x_1 + x_4$ in the Supplementary Figure S26.
Reconstructing the corresponding attractor would enable the assessment of key topological features associated with epilepsy dynamics, and help predict the spatio-temporal diversity of seizure propagation \cite{proix2018predicting}.

\begin{figure}[ht!]
    \centering
    \includegraphics[width=1\linewidth]{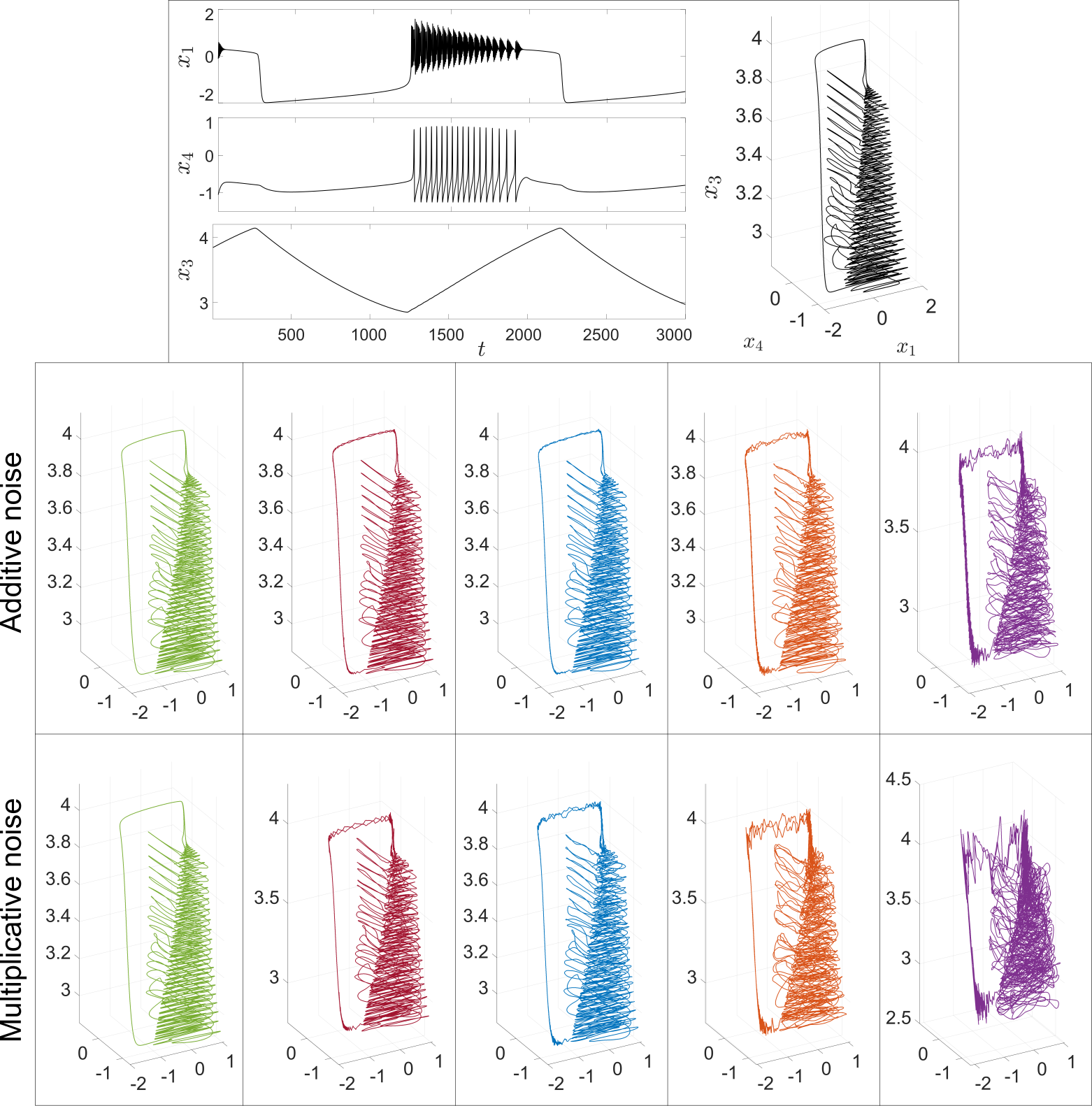}
    \caption{\footnotesize \textbf{The Differentiator enables de-noising and faithful attractor reconstruction for the Epileptor system affected by various types of additive and multiplicative noise, in the state-space coordinates.} Reconstructed attractor of the Epileptor model \eqref{eq:Epileptor} from measurements of the signals $x_1$, $x_4$ and $x_3$ affected by either \textit{additive} or \textit{multiplicative} noise, obtained through the HD-based method. The Differentiator parameters are set as $n_d=2$, $n_f=10$, $L_0=50$. \textbf{Rows, from top to bottom:} noise-free signals $x_1$, $x_4$ and $x_3$ to be measured, along with the true attractor of the system (black) in the $\{x_1, x_3, x_4\}$ sub-space; attractor reconstruction with additive noise; attractor reconstruction with multiplicative noise.
    \textbf{Columns, from left to right:} both in the case of additive and of multiplicative noise, reconstruction of the attractor (in the same space) from measurements corrupted by noise $\eta_H(t)$ as in \eqref{eq:harmonic_noise} (green); $\eta_U(t)$ as in \eqref{eq:unbounded_noise} (red); $\eta_1(t) \sim \mathcal{N}(0, 0.01)$ (blue); $\eta_2(t) \sim \mathcal{N}(0, 0.1)$ (orange); $\eta_3(t) \sim \mathcal{N}(0, 1)$ (violet).}
    \label{fig:Epileptor attractor reconstruction}
\end{figure}

Figure~\ref{fig:Epileptor attractor reconstruction} shows the attractor reconstruction from noisy measurements of the signals $x_1$, $x_3$ and $x_4$ in the $\{x_1, x_3, x_4\}$ sub-space, obtained with our proposed HD-based method. Comparing the reconstruction with the true attractor leads to errors $\tilde E_R$ as in \eqref{eq:E_R_orig_coord} that are comparable with the errors $E_R$ as in \eqref{eq:E_R_time_delay} obtained for the Lorenz system, even though the model is more complex than the HR model: in fact, in this case, we reconstruct the attractor in the $\{x_1, x_3, x_4\}$ sub-space from measurements of all the three sub-space variables, without employing their derivatives, while for the HR model we only measure one variable and we exploit its derivative to estimate the other variable of the sub-space where the attractor is reconstructed; for additive noise, this results in better theoretically provable bounds on the estimation error (see Theorem 3 in the Supplementary Material). This example, thus, highlights the highly beneficial effect of having access to measurements of more variables.
For all noise types, the relative errors are, on average over all data points, less than 10\%. Precise attractor reconstruction and noise filtering are ensured in less than one minute: the computation time is larger than for the Lorenz system due to the larger number of simulated data points, which are 3,000,001 for the Epileptor model.

Quantitative information about the distributions of errors and their average statistics is provided in Section S4 of the Supplementary Material, along with the HD-based reconstruction of the attractor in the $\{\bar{y},\dot{\bar{y}}\}$ space from measurements of the seizure-like event signal $\bar{y}(t)=x_1(t)+x_4(t)$.

\section{Reconstructing Unknown Attractors from Empirical Data}
\label{sec:empirical_data}

Faithful attractor reconstruction is fundamental to validate
%Observing the emergence of clear attractors provides a step forward, towards the validation of
models for brain dynamics, whose evolution is based on complex attractors \cite{breakspear2017dynamic,freyer2009bistability} and on switches among them, and hence to inform the development of interpretable and effective algorithms to detect and predict transitions from healthy to disease states \cite{lucas2024topological}. As a key example, unravelling the underlying dynamics of epileptic seizures would help develop accurate and interpretable methods to predict seizure events and improve epilepsy management, by providing early warnings for patients and triggering interventions \cite{kuhlmann2018epilepsyecosystem}. The analysis of the topological properties of dynamical attractors can help discover universal routes to
epilepsy, verify computational methods to benefit patients, foster new methods for early warning, and inform intervention and control strategies.
In the previous section, we have demonstrated that our HD-based methodology effectively reconstructs from noisy measurements the attractor of the Epileptor model \cite{jirsa2014nature}, widely used to simulate seizure-like events, and identifies the underlying dynamics in a model-free fashion, by properly de-noising corrupted signals.
We now showcase that the Differentiator can reconstruct the core dynamics of real brain signals, and identify the key topological characteristics of attractors reconstructed from noisy experimental data.

In particular, we apply the HD-based attractor reconstruction methodology to an empirical dataset of Local Field Potential (LFP) recordings, obtained from measuring the brain activity of Zebrafish (\textit{Danio rerio}) larvae from the genetic strain KCNQ5 Loss-of-Function model kcnq5a$^{\text sa9563}$; details about handling, ethical statements and husbandry can be found in \cite{moein_casian_2018,martins_seizure-induced_2023}. The dataset is publicly accessible under doi \texttt{10.17881/8hpj-0f07} and from \cite{lucas2024topological}. The data were generated by controlled experiments on genetically mutated fish, which experience several events of epileptic seizures during the recording, and are thus particularly suitable for controlled tests. Using these data allows us to apply the HD-based methodology on curated time series, with known sampling frequencies, and showing different dynamical regimes: normal activity (``background'') and ictal events (``seizure''). The data are provided both as unlabelled and unprocessed time series, with sampling rate $100,000$ Hz, over more than $30$ minutes, or as downsampled, segmented and labelled (as either Background or Seizure) events, each spanning several seconds, with sampling rate $2,000$ Hz. We select a random sample to perform our preliminary tests.

\begin{figure}[ht!]
    \centering
    \includegraphics[width=1\linewidth]{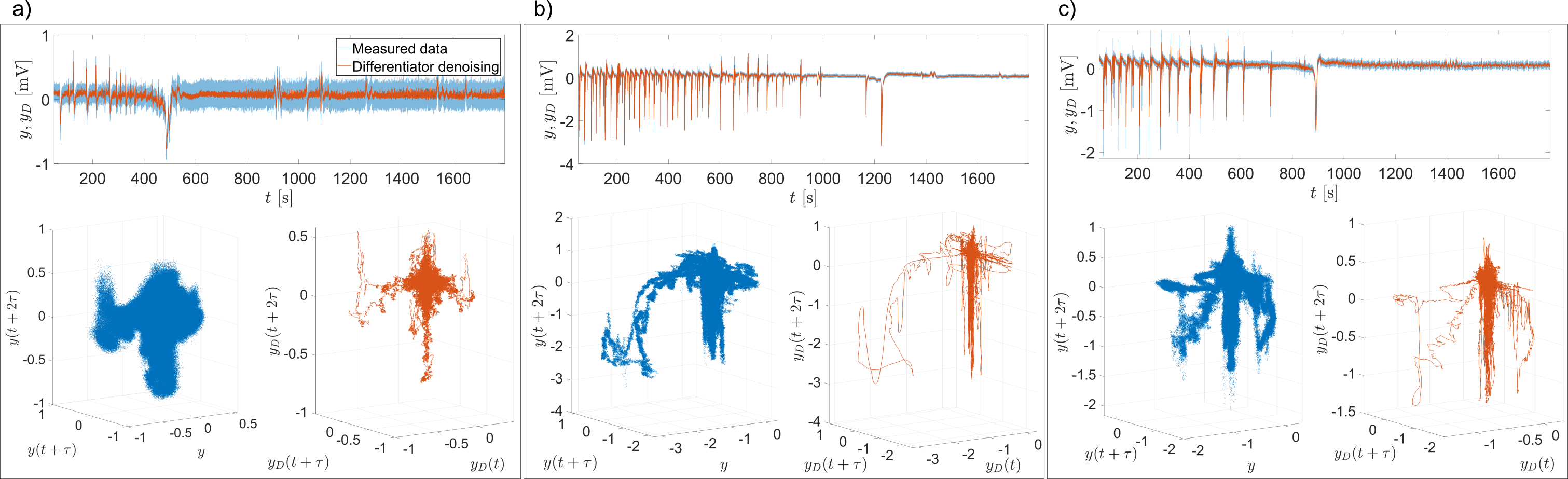}
    \caption{\footnotesize \textbf{The Differentiator enables signal de-noising and attractor reconstruction from an empirical dataset of Local Field Potential recordings of the brain activity of Zebrafish larvae, and reveals key topological structures that we hypothesise are associated with Background activity (central bulks) and with Seizure events (semi-loops).}
    \textbf{Top:} Unlabelled and unprocessed Local Field Potential recordings at $100,000$ Hz of the brain activity of Zebrafish larvae (blue) and the corresponding de-noised signal provided by the Differentiator with parameters $n_d=3$, $n_f=9$, $L_0=750$ (orange). \textbf{Bottom:} Three-dimensional Takens embedding obtained from the noisy measured data (blue) and from the de-noised signal provided by the Differentiator (orange). The computed time-delays are: \textbf{a)} $\tau=180$ $[s]$; \textbf{b)} $\tau=576$ $[s]$; and \textbf{c)} $\tau=398$ $[s]$.}
    \label{fig:zebra_fish_full_data}
\end{figure}

Figure~\ref{fig:zebra_fish_full_data} shows the reconstructed attractor (displayed in three-dimensional time-delay coordinates for clarity of visualisation) for the full $100,000$ Hz time series, as well as the corresponding de-noised signal. The computation time to process the $180,000,000$ data points is between $23$ and $24$ minutes. The Differentiator drastically suppresses most of the measurement noise, even in case of sudden spikes (which, according to neurologists, are possibly associated with fish movements). The attractor reconstruction is performed using time-delay embedding, with $\tau$ chosen as described in Section~\ref{sec:methods}. Differently from the direct embedding using noisy measurements, reconstructing the attractor after de-noising the signal with the Differentiator allows us to recognise key topological structures that are otherwise obscured by the noise: a central ``bulk'' is visible along with some elongation, as well as semi-loops similar to that observed in the Epileptor attractor in Figure~\ref{fig:Epileptor attractor reconstruction}. Pre-processing the data using the Differentiator thus bears the promise to retain crucial topological information that can improve the resolution and results of further studies, such as the measurement of persistent homology features across species and genetic conditions.

\begin{figure}[ht!]
    \centering
    \includegraphics[width=1\linewidth]{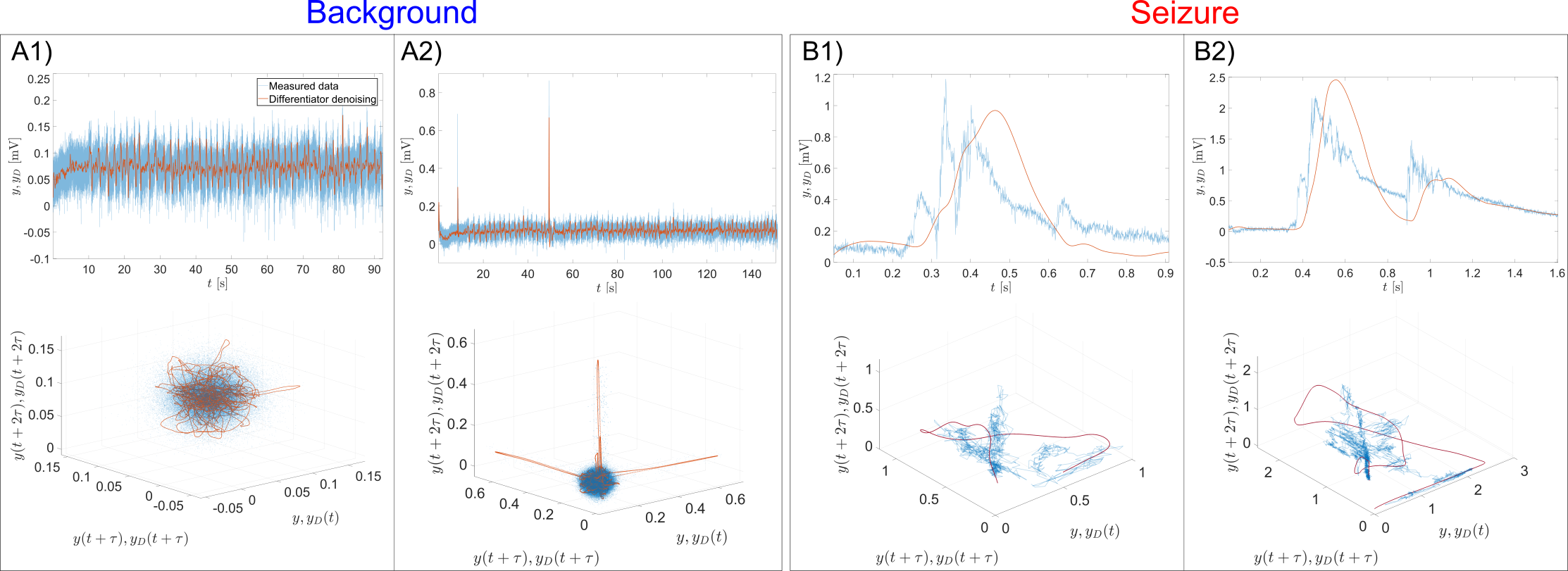}
    \caption{\footnotesize \textbf{The Differentiator enables signal de-noising and attractor reconstruction from segmented Local Field Potential empirical recordings of the brain activity of Zebrafish larvae, and reveals distinct key topological structures: central bulks in segments labelled as Background activity and semi-loops in segments labelled as Seizure events.} \textbf{Top:} Local Field Potential recordings at $2,000$ Hz of the brain activity of Zebrafish larvae (blue) and the corresponding de-noised signal provided by the Differentiator with parameters $n_d=3$, $n_f=9$, $L_0=750$ (orange). \textbf{Bottom:} Three-dimensional Takens embedding obtained from the noisy measured data (blue) and from the de-noised signal provided by the Differentiator (orange), for data labelled as Background (left) and Seizure (right). The computed time-delays are: \textbf{A1)} $\tau=26.4$ $[s]$; \textbf{A2)} $\tau=10.1$ $[s]$; \textbf{B1)} $\tau=0.159$ $[s]$; \textbf{B2)} $\tau=0.167$ $[s]$.}
    \label{fig:zebra_fish_segmented_data}
\end{figure}

Furthermore, Figure~\ref{fig:zebra_fish_segmented_data} shows the signal de-noising and attractor reconstruction (in three-dimensional time-delay coordinates as above) for the segmented $2,000$ Hz time series of background and seizure activity. These are different events occurred on a single fish, during the recordings (other events for the same fish are shown in the Supplementary Figure S27). As before, the Differentiator suppresses most of the measurement noise, ideally unveiling the core dynamics and highlighting the key topological differences between the two types of events. Background events are characterised by an agglomerate of points that, except for punctual and sparse spikes, are akin to typical attractors of semi-random signals; this is consistent with the neurological hypothesis that the global brain dynamics is overall incoherent \cite{bullock1995temporal, ghosh2008noise}. For seizure events, the Differentiator highlights a clear structure that evolves as a semi-loop and suggests some form of pattern or coherence; this is in line with the neurological hypothesis that seizures emerge from spontaneous alignment of neural firing \cite{jiruska2013synchronization, detti2020eeg}. Moreover, these semi-loop structures may be linked to the loops observed in Figure~\ref{fig:zebra_fish_full_data}, which are captured by dynamical models such as the Epileptor. If our hypothesis were confirmed, the Differentiator would emerge as the tool of choice to help verify models and topological characteristics of seizure events, thus improving their detection and prediction. The considered dataset does not provide a mapping between the segmented time series with labelling of conditions and the full unsegmented time series; hence, we cannot fully test this interesting hypothesis, which is left for future investigation.

Overall, our preliminary results demonstrate the impressive capability of the Differentiator to process and de-noise empirical signals, and to reconstruct cleaned-up attractors, thus highlighting their key topological features that would otherwise be obscured by unknown noise. In turn, unveiling these features enables the development of new hypotheses, the verification of dynamical models from data, and a deeper subsequent analysis.

\section{Discussion}
\label{sec:discussion}

We have proposed the use of the Homogeneous filtering Differentiator \cite{levant2003higher,levant2017sliding,levant2020robust,hanan2021low} as a powerful, computationally efficient and interpretable methodology to provide de-noised estimates of noisy signals and of their time derivatives, as well as to reconstruct dynamic attractors of complex nonlinear systems from noisy measurements, thus casting light onto their fundamental properties and qualitative behaviours.
The Differentiator can effectively handle a wide variety of noise types, affecting the signal both in an additive and in a multiplicative fashion, and comes with rigorous theoretical guarantees in the case of additive noise. In addition, the Differentiator parameters are few and easy to tune without any knowledge about the system under study, and the estimates provided by the Differentiator converge to the de-noised signal and its derivative in finite time (rather than asymptotically).
Using well-known and challenging dynamical systems endowed with complex attractors -- the Lorenz system, the Hindmarsh-Rose model and the Epileptor model -- we have conducted a systematic performance analysis and comparison with existing methods. Our numerical results have shown that the Differentiator enables faithful and efficient attractor reconstruction, both in the original state-space coordinates and in the time-delay embedding coordinates \cite{Takens1981}, surpassing existing approaches both in terms of accuracy and in terms of computational time.

In particular, we have compared our proposed Differentiator-based methodology with attractor de-noising via the Schreiber-Grassberger algorithm \cite{SchreiberGrassberger1991,grassberger1993noise}, which is to date the most frequently used and recommended technique across fields. We recall that alternative approaches for noise reduction schemes exist, although they are not commonly used \cite{sardy2001robust}; nonetheless, they still suffer from similar problems as the Schreiber-Grassberger method, namely limited theoretical guarantees, scope limited to specific noise types and to noise affecting the signal in an additive fashion, and long computation times that make the methods computationally prohibitive when the number of data points is large. Moreover, they only allow for time delay embedding and they cannot enable differential embedding or direct reconstruction of the attractor in the state space coordinates, which requires the ability to estimate not just the measured signal, but also its time derivatives. %: for instance, \cite{sardy2001robust} proposes a wavelet-based de-noising technique for signals corrupted by additive Gaussian and Student's $t$ noises, but the choice of the algorithm parameters is mainly heuristic and based on statistical considerations of the expected noise.
The Differentiator-based approach overcomes all these limitations, and allows to faithfully reconstruct dynamic attractors of complex systems in seconds instead of several hours as in the Schreiber-Grassberger algorithm. We remind that the use of the HD requires the selection of parameter $L_0$, which we have currently fixed using post-hoc analysis; determining effective heuristics to set $L_0$ a priori is a challenging avenue for future work.

Attractor reconstruction for important models in neurosciences, such as the Hindmarsh-Rose model, has allowed us to distinguish between the different regimes that characterise the system behaviour, and to gain additional insight into their main features, in terms of dynamic attractors.
Further studies may explore the ability of the Differentiator to shed light onto neural computation properties, for instance by studying the types of bifurcations associated with the model transitioning from one regime to another, and hence with the excitability of neurons and its neurocomputational properties \cite[Chapter 7]{izhikevich2007dynamical}. %\GG{For instance, the square-wave bursting regime (third row in Figures~\ref{fig:HR additive noises} and~\ref{fig:HR multiplicative noises}) is known to show the signatures of a homoclinic bifurcation of a saddle equilibrium (each spike in the time evolution of the signal $x_1$ is large and the spiking frequency decreases as the bifurcation point is approached), while the plateau-bursting regime (fourth row in Figures~\ref{fig:HR additive noises} and \ref{fig:HR multiplicative noises}) shows the qualitative footprints of a supercritical Andronov-Hopf bifurcation (the amplitude of the spikes in the time evolution of the signal $x_1$ approaches zero as the bifurcation point is approached). \textbf{[Shall we include this? If so, clarify! Bifurcations are associated with transitions from one regime to the other, not with regimes. Add references to the papers where these associations have been put forth!]}}

Moreover, we have tested the proposed Differentiator-based method for attractor reconstruction on real noisy measurement data from empirical Local Field Potential recordings obtained from measurements of the brain activity of Zebrafish larvae. The results showcase an impressive ability of the methodology to de-noise the measured signals and the corresponding reconstructed attractors, thus revealing characteristic features that discriminate between background brain activity and seizure events, and allowing us to formulate promising hypotheses. In this sense, the HD is a powerful processing methodology for a broader analysis of experimental data generated by complex dynamics, which paves the way to future studies leveraging attractor reconstruction results. It is unfortunate that, due to the pre-processing that affects the available dataset, it is not possible to map the segmented data at a low frequency (labelled as Background and Seizure by experts) onto the original complete time series at a high frequency, so as to verify our detection and prediction results, and already test our hypotheses.
The analysis of an additional and more complete dataset, which would allow us to test the hypotheses discussed in the Results section, is therefore left as a future research direction.

%Also remind that we have chosen the metaparameters a posteriori. ---> Just L0... perhaps not worth stressing?
%Methods to identify the best ones before the processing are also demanded for future works.

%using the denoising method as a preprocessing step, we expect future works to significantly improve the results of learning algorithms, dynamic characterisation etc. ---> I would not focus on this, it seems like we are "selling" the Differentiator but we are not the creators, just users... what we are proposing here is using it for attractor reconstruction

\section*{Acknowledgments}
The authors thank prof. Alexander Skupin for direct access to Zebrafish data and discussions about the interpretation of their analysis.

\section*{CRediT authorship contribution statement}
\textbf{U.S.:} conceptualisation, formal analysis, investigation, methodology, software, validation, visualisation, writing—original draft, writing—review and editing.

\textbf{D.P.:} conceptualisation, formal analysis, methodology, validation, visualisation, writing—original draft, writing—review and editing.

\textbf{R.K.:} conceptualisation, formal analysis, methodology, validation, writing—original draft, writing—review and editing.

\textbf{G.G.:} conceptualisation, formal analysis, funding acquisition, methodology, project administration, supervision, writing—original draft, writing—review and editing.

\section*{Code availability}
The \textsc{Matlab} code used to generate our results as well as the Supplementary Material are publicly available on GitHub at
{\texttt{https://github.com/Uros-S/DiffAttractReconstruct}}

\section*{Data availability}
All the data used to support the findings of this study are included in the article.

\section*{Funding}
Work supported by the European Union through the ERC INSPIRE grant (project number 101076926). Views and opinions expressed are however those of the authors only and do not necessarily reflect those of the European Union or the European Research Council Executive Agency. Neither the European Union nor the European Research Council Executive Agency can be held responsible for them.

\section*{Declaration of competing interest}
The authors declare no competing interests.

%\appendix

%\section{Sample Appendix Section}
%Lorem ipsum dolor sit amet, consectetur adipiscing elit, sed do eiusmod tempor section~\ref{sec:sample1} incididunt ut labore et dolore magna aliqua. Ut enim ad minim veniam, quis nostrud exercitation ullamco laboris nisi ut aliquip ex ea commodo consequat. Duis aute irure dolor in reprehenderit in voluptate velit esse cillum dolore eu fugiat nulla pariatur. Excepteur sint occaecat cupidatat non proident, sunt in culpa qui officia deserunt mollit anim id est laborum.


\restoregeometry

 \bibliographystyle{elsarticle-num} 
 \bibliography{cas-refs}





% \end{thebibliography}

\end{document}